\documentclass[iop]{emulateapj}   

\usepackage{comment}
\usepackage{hyperref}
\usepackage[flushleft]{threeparttable}

\usepackage{amsmath}
\usepackage{epstopdf}

\shorttitle{}
\shortauthors{}

\begin{document}

\title{Hidden in plain sight: A massive, dusty starburst in a galaxy protocluster at $z=5.7$ in the COSMOS field}

\author{Riccardo Pavesi\altaffilmark{1}$^\dagger$, Dominik A. Riechers\altaffilmark{1}, Chelsea E. Sharon\altaffilmark{2}, Vernesa Smol\v{c}i\'c\altaffilmark{3}, Andreas L. Faisst\altaffilmark{4},\\ Eva Schinnerer\altaffilmark{5}, Christopher L. Carilli\altaffilmark{6,7}, Peter L. Capak\altaffilmark{4}, Nick Scoville\altaffilmark{8}, Gordon J. Stacey\altaffilmark{1}}

\affil{$^1$Department of Astronomy, Cornell University, Space Sciences
Building, Ithaca, NY 14853, USA \\$^2$Department of Physics \& Astronomy, McMaster University,   1280 Main Street West, Hamilton, ON L85-4M1, Canada\\
$^3$University of Zagreb, Physics Department, Bijeni\v{c}ka cesta 32, 10002 Zagreb, Croatia\\ $^4$Infrared Processing and Analysis Center, California Institute of Technology, Pasadena, CA 91125, USA\\ $^5$Max Planck Institute for Astronomy, K\"onigstuhl 17, D-69117, Heidelberg, Germany\\$^6$National Radio Astronomy Observatory, PO Box O, Socorro,
NM 87801, USA\\ $^7$Cavendish Astrophysics Group, University of Cambridge,
Cambridge, CB3 0HE, UK\\$^8$Astronomy Department, California Institute of Technology, MC 249-17, 1200 East California Boulevard, Pasadena, CA
91125, USA\\  }
     \email{$^\dagger$rp462@cornell.edu}


\begin{abstract}
We report the serendipitous discovery of a dusty, starbursting galaxy at $z=5.667$ (hereafter called CRLE) in close physical association with the ``normal" main-sequence galaxy HZ10 at $z=5.654$. CRLE was identified by detection of  [C{\sc ii}], [N{\sc ii}] and CO(2--1) line emission,  making it the highest redshift, most luminous starburst in the COSMOS field. This massive, dusty galaxy appears to be forming stars at a rate of at least 1500$\,M_\odot$ yr$^{-1}$ in a compact region only $\sim3$ kpc in diameter. The dynamical  and  dust emission properties of CRLE suggest an ongoing merger driving the starburst, in a potentially intermediate stage relative to other known dusty galaxies at the same epoch. The ratio of [C{\sc ii}] to [N{\sc ii}] may suggest that an important ($\sim15\%$) contribution to the [C{\sc ii}] emission comes from a diffuse ionized gas component, which could be more extended than the dense, starbursting gas. CRLE appears to be located in a significant  galaxy overdensity at the same redshift,  potentially associated with a large-scale cosmic structure recently identified in a Lyman Alpha Emitter survey. This overdensity suggests that CRLE and HZ10 reside in a protocluster environment, offering the tantalizing opportunity to study the effect of a massive starburst on protocluster star formation. Our findings support the interpretation that a significant fraction of the earliest galaxy formation may occur from the inside out, within the central regions of the most massive halos, while rapidly evolving into the massive galaxy clusters observed in the local Universe.
\end{abstract}

\section{Introduction}
While a significant fraction of the visible sky is covered by moderate-redshift, low-mass galaxies, the view of dust-obscured star formation in the sub-millimeter sky preferentially selects high-mass, dusty star-forming galaxies (DSFGs) at high redshift. This complementary view of the Universe can provide unique insights into galaxy formation processes (e.g., \citealt{BlainReview,CaseyReview}).
While the general population of DSFGs was  found to predominantly occupy the peak epoch of cosmic star formation at $z=2-3$, a significant  tail of higher redshift, and often brighter, examples appears to already  be  in place at $z>5$ (e.g., \citealt{Riechers10a,RiechersHFLS3,Riechers17,Walter12,Weiss13,Strandet16,Strandet2017}). 
These submillimeter-selected galaxies  in the early Universe are often extreme ``hyper-starbursts", reaching infrared luminosities of $L_{\rm IR}>10^{13}\,L_\odot$, and star formation rates (SFRs) exceeding 1000$\,M_\odot\,$yr$^{-1}$ within small spatial regions of a few kiloparsecs in diameter. They are likely the result of major mergers and/or extreme gas accretion events, which may only be possible in the highest density regions of the early Universe (e.g., \citealt{Riechers_GN19,Riechers17,Ivison_GN19,Ivison13, Capak11,Oteo2016, Marrone2017}).

Recent single-dish studies of the evolution of the sub-millimeter luminosity function have suggested that, while the space density of DSFGs may significantly decrease at $z>3$, the knee of the luminosity function appears to shift to higher infrared luminosities, perhaps capturing the high redshift tail of ``titans" (e.g., \citealt{Koprowski17}).
While the $z>4$ infrared luminosity function is far from accurately constrained, and the diversity of processes that may produce dusty hyper-starbursts are not completely understood (e.g., \citealt{Narayanan15}), these results may not be completely surprising in the context of  recent theoretical work exploring the role of protocluster star formation (e.g., \citealt{Chiang2013,Chiang2017}). In particular, simulations suggest that a large fraction of  early Universe star formation may have taken place in the densest regions, traced by the most massive halos, in an inside-out fashion. That is, a significant fraction of galaxy formation may have started in protocluster cores where the gas densities and the galaxy merger probability would have been highest, implying that this star formation may have been bursty and highly dust-obscured, as seen in high-redshift DSFGs.
The importance of the connection between DSFGs and galaxy overdensities has been known for some time (e.g., \citealt{Aravena2010}). Even the first $z>5$ DSFG known, AzTEC-3 at $z=5.3$, was quickly recognized to be located near the center of a rich galaxy protocluster \citep{Capak11,Riechers14}.  Recent studies have explored the connection between DSFGs and galaxy overdensities in detail, finding evidence for a strong relationship, potentially getting stronger with redshift toward the early Universe (e.g., \citealt{Vernesa_overdens,Lewis2017}). Two notable case studies  of clustered, dusty galaxy formation are the extremely rich protocluster of dusty star forming galaxies identified at $z=4$ by \cite{Oteo17}, and the surprisingly high incidence (4/25) of very close association ($<$600 kpc) of dusty star forming galaxies to $z>6$ quasars \citep{Decarli2017}.  The former case shows that several protocluster members may be experiencing a dusty starburst phase simultaneously, perhaps triggered by a massive gas flow. Both cases show that  active galactic nuclei (AGN) or massive starburst galaxies may be found in association and can be serendipitously discovered in sensitive Atacama Large (sub-)Millimeter Array (ALMA) observations.

Here we describe the serendipitous discovery of the brightest and highest redshift dusty starburst in COSMOS \citep{Nick_COSMOS} at J2000 10$^h$0$^m$59$^s$.2, 1$^\circ$33$^\prime$6.6$^{\prime\prime}$, which we identified in ALMA observations of atomic fine-structure lines in a close-by galaxy, HZ10. This new DSFG is located only 13$^{\prime\prime}$ ($\sim77$ kpc) away from HZ10, which is an above average dusty but ``normal" (i.e., $\sim L^*_{\rm UV}$) galaxy at $z=5.654$ \citep{C15,Pavesi16}.
HZ10 is a Lyman Alpha Emitter (LAE; \citealt{Murayama07}) and a Lyman Break Galaxy (LBG) which was selected for [C{\sc ii}] $158\,\mu$m and dust emission observations with ALMA based on its strong ultraviolet absorption features \citep{C15}.
We subsequently  followed up HZ10 in [N{\sc ii}] 205$\,\mu$m emission, and CO(2--1) (R. Pavesi et al., in prep.). These observations provide additional support to the interpretation that HZ10 appears to show a more ``mature" and metal-rich inter-stellar medium (ISM) than other ``typical" massive galaxies at $z>5$ \citep{Pavesi16, Faisst17}. 
The new  DSFG is also detected in  {\it Herschel} observations. Its ``red" color in the  SPIRE $250-500\,\mu$m bands is consistent with its high redshift.
We refer to this galaxy as COSMOS (FIR-)Red Line Emitter (CRLE, read ``curly") in the following, after the methods that lead to its discovery.

In Section~2, we describe the spectroscopic observations that allowed the identification of CRLE as a dusty starburst at $z\sim5.7$. In Section~3, we present the results of our line measurements and discuss their implications. In Section~4, we present the continuum measurements and discuss the dust properties of CRLE and the evidence for a galaxy merger. In Section~5, we analyze public galaxy catalogs to identify and characterize a galaxy overdensity around CRLE, providing evidence for a galaxy protocluster. We conclude in Section~6. An obstacle to any previous identification of this source may have been the presence of an unrelated foreground ($z\sim0.3$), low-mass spiral galaxy, which covers the optical counterpart of CRLE\footnote{For this reason, and for its high sub-mm flux, CRLE may be considered ``hidden in plain sight".}, and therefore prevents detection in the visible and near-IR (NIR) bands.   In Appendix A, we detail our constraints on the potential contribution of gravitational lensing from the foreground galaxy to the observed properties of CRLE, finding that lensing may be negligible. Appendix B details our attempt to separate the emission from CRLE from that of the foreground galaxy at observed-frame NIR wavelengths. Appendix C contains a description of the {\sc uv}-space dynamical modeling technique we utilized to investigate the [C{\sc ii}] observations from CRLE. In this work, we adopt  a Chabrier IMF and a flat, $\Lambda$CDM cosmology with $H_0=70\,$km s$^{-1}$ Mpc$^{-1}$ and $\Omega_{\rm M}=0.3$.  Quoted lengths are proper sizes unless otherwise specified (for comoving distances).
 



\section{Observations}

\subsection{ALMA observations of {\rm [C{\sc ii}]} and {\rm [N{\sc ii}]}}
The ALMA data containing the [C{\sc ii}] line for CRLE were first presented by \cite{C15} as Cycle 1 observations of HZ10. A subsequent reprocessing of those data, together with a description of part of the Cycle 3 [N{\sc ii}] observations, was described in detail by \cite{Pavesi16}.

The Cycle-1 observations of [C{\sc ii}] were taken on 2013 November 16 in band 7 as part of a larger project (ID: 2012.1.00523.S; \citealt{C15}).
The pointing resulted in $56\,$min on source time with 25 usable antennas. The  primary beam attenuation factor at the position of CRLE is $\sim3$.
The correlator was set up to target the expected frequency of the [C{\sc ii}] line in HZ10 at 284.835 GHz,  and to provide continuous coverage of the continuum emission in adjacent spectral windows (centered $\sim2$, 12, and 14 GHz above) with channels of $15.6\,$MHz width in Time Division Mode (TDM). 
The synthesized beam size is approximately $0^{\prime\prime}.6\times0^{\prime\prime}.5$, when adopting natural baseline weighting. We refer to \cite{Pavesi16} for a complete description of the observations and the imaging product.

 Cycle 3 observations of [N{\sc ii}] 205$\,\mu$m targeting HZ10 and also covering CRLE were taken on 2016 January 1 and 5 in band 6, as part of two separate programs (2015.1.00928.S and 2015.1.00388.S; PIs: Pavesi and Lu, respectively) with one track each in a compact configuration (max. baseline $\sim300\,$m).
The two sets of observations resulted in 50 min, and  35 min on source with $\sim$41--45 usable $12\,$m antennas under good weather conditions at 1.3 mm.
The first set of observations was previously described by \cite{Pavesi16}. 
For the second set of observations, the nearby radio quasar J0948+0022 was observed regularly for amplitude and phase gain calibration, J1058+0133 was observed for bandpass calibration, and Callisto was used for flux calibration. We estimate the overall accuracy of the flux calibration to be within $\sim10\%$.
The correlator was set up to cover two spectral windows of 1.875 GHz bandwidth each at $15.6\,$MHz ($\sim20\,$km s$^{-1}$) resolution (dual polarization) in TDM, in each sideband.

We used the Common Astronomy Software Application ({\sc casa}) version 4.5 for data reduction and analysis.  We combined data from all observations and produced all images with the {\sc clean} algorithm, using natural weighting for maximal sensitivity.
 Imaging these [N{\sc ii}] data results in a synthesized beam size of $1.6^{\prime\prime} \times 1.2^{\prime\prime}$ at the redshifted  [N{\sc ii}] frequency of CRLE and in the continuum map. The rms noise in the phase center is $\sim0.14\,$mJy beam$^{-1}$ in a $44\,$km s$^{-1}$ wide channel.  The final rms noise when averaging over the line-free spectral windows (i.e. over a total $7.1\,$GHz of bandwidth) is $\sim 18\, \mu$Jy beam$^{-1}$. The  primary beam attenuation factor at the position of CRLE is $\sim1.8$.

\begin{figure*}[t]
\centering{
 \includegraphics[width=\textwidth]{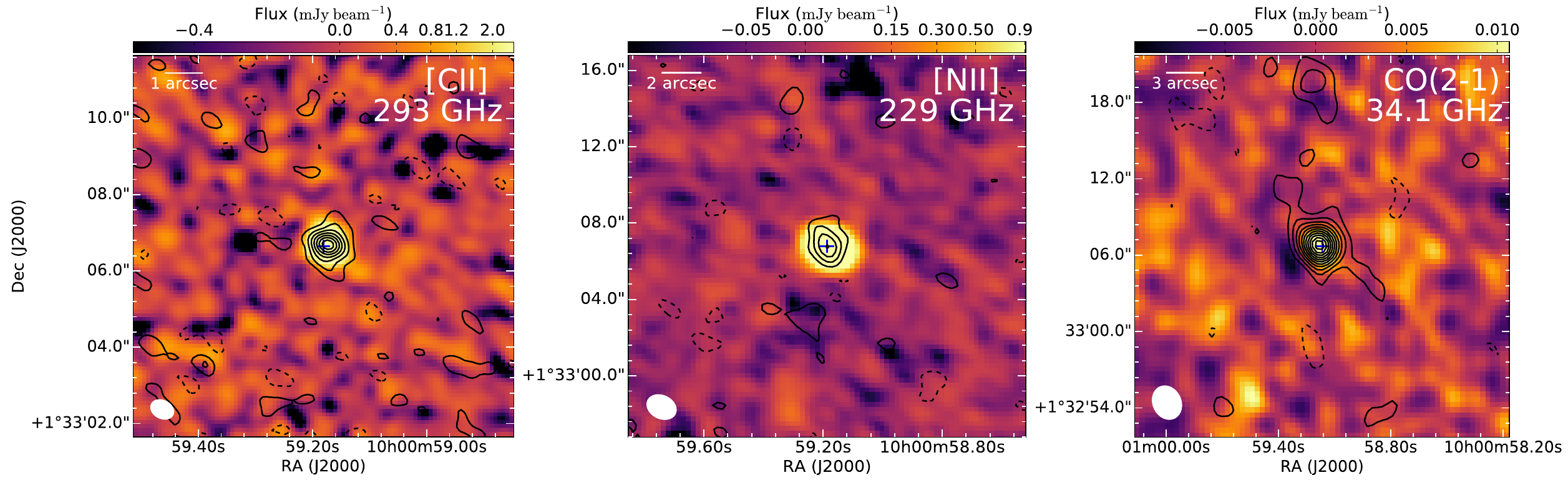}
}
\caption{Continuum and integrated line maps for CRLE. The contours show the [C{\sc ii}] (left, observed with  ALMA), [N{\sc ii}] (middle, observed with ALMA), and CO(2--1) (right,  observed with the VLA) emission, while the background images show the continuum emission from the corresponding observations (frequencies are shown in the observed-frame). Blue crosses indicate the positions of the continuum peaks. The  synthesized beam size is shown in the bottom left corner of each panel.  Contours are multiples of $2\sigma$, starting at $\pm2\sigma$.
The [C{\sc ii}] line, $158\,\mu$m continuum (corresponding to observed-frame 293 GHz), and $205\,\mu$m continuum (corresponding to observed-frame 229 GHz) emission are resolved. The [N{\sc ii}] line emission is marginally resolved.}
\label{fig:mom_cont}
\end{figure*}

\subsection{VLA observations of {\rm CO(2--1)}}
We observed the CO(2--1) transition line in CRLE ($\nu_{\rm rest}$=230.538 GHz, redshifted to $\sim$34.58$\,{\rm GHz}$ at $z\sim5.667$), using NSF's Karl G. Jansky Very Large Array (VLA) in Ka band (project ID: 17A-011, PI: Pavesi). Observations were carried out between 2017 March 4 and April 6, with 27 antennas in the most compact array configuration (D; max. baseline $\sim950\,$m) under good to moderate weather conditions at 35 GHz (precipitable water vapor columns of 3--$6\,$mm) for eight sessions.
The pointing for these observations was centered on HZ10.  The primary beam attenuation factor at the position of CRLE is  $\sim$1.08. The combination of all data results in a total on-source time of $19.8\,$hr.

The nearby radio quasar J1041+0610 was observed regularly for amplitude and phase gain calibration. Also, 3C286 was observed for bandpass and flux calibration. We estimate the overall accuracy of the flux calibration to be within $\sim10\%$, since the phase calibrator flux was measured to be constant within $<5\%$ across all sessions.

The VLA correlator was utilized in 8-bit mode for maximum sensitivity.
In the first three sessions, the correlator was set up with two intermediate-frequency bands (IFs) of eight, 128 MHz-wide spectral windows each, centering one IF on the redshifted CO(2--1) line frequency in HZ10 and the other contiguously below to provide additional continuum sensitivity.
The lower IF was moved in the remaining five sessions, by centering it on the CO(2--1) line in CRLE (which otherwise fell onto a sub-band gap).
While these overlapping sidebands limit the simultaneous bandwidth and hence the continuum sensitivity, they provide uninterrupted coverage of the CO(2--1) line in both galaxies.
The channels in all spectral windows were chosen to provide 1 MHz ($\sim9\,$km s$^{-1}$) resolution (dual polarization) in each IF, to obtain a simultaneous bandwidth of 2.048 GHz for the first three sessions and 1.349 GHz for the remaining five sessions.
Because of overlapping spectral coverage, the measurements in the two IFs are not independent. Therefore, we never combine their data, but rather only use the line IF for the analysis of CRLE.

We used {\sc casa} version 4.7  for data reduction and analysis. We calibrated the visibilities using the scripted VLA pipeline, supplemented by manual flagging through inspection of  standard visibility plots.
We combined data from all observations and imaged them  with the {\sc clean} algorithm, using natural weighting for maximal sensitivity.
The imaging of the CO(2--1) data results in a synthesized beam size of $2.7^{\prime\prime} \times 2.3^{\prime\prime}$ at the redshifted  CO(2--1) frequency and in the continuum map. The rms noise in the phase center is $\sim45\,\mu$Jy beam$^{-1}$ in a $35\,$km s$^{-1}$ wide channel.  The final rms noise when averaging over the line-free  $1.92\,$GHz of bandwidth is $\sim 2.7\, \mu$Jy beam$^{-1}$. 


\begin{table*}[t]
\caption[]{Measured line properties of CRLE.}
\label{table_lines}
\centering
\begin{tabular}{l c c c c }
\hline 
&[C{\sc ii}] &[N{\sc ii}] & CO(2--1) \\
\hline
Component 1\\
$\nu_{{\rm obs}}$(GHz) &$284.892\pm0.013$ &219.0298$^a$&$34.5604\pm0.0017$\\
$S_{\rm peak}$(mJy) &$16\pm3$  &$1.2\pm0.2$ &$0.49\pm0.04$ \\
FWHM (km s$^{-1}$) &$280\pm40$&280$^a$& $290\pm30$\\
$I$ (Jy km s$^{-1}$) & $4.5\pm0.9$&$0.33\pm0.07$ &$0.14\pm0.02$ \\
\hline
Component 2\\
$\nu_{{\rm obs}}$(GHz) &$285.27\pm0.07$ & 219.3204$^a$& $34.613\pm0.004$ \\
$S_{\rm peak}$(mJy) &  $9.6\pm0.9$&$0.65\pm0.13$&$0.25\pm0.03$ \\
FWHM (km s$^{-1}$) & $610\pm140$&610$^a$&$420\pm90$\\
$I$ (Jy km s$^{-1}$) & $6.3\pm1.5$&$0.40\pm0.13$ &$0.11\pm0.03$\\
\hline
Total\\
Mean redshift& $5.6666\pm0.0008$\\
$I$ (Jy km s$^{-1}$) & $10.8\pm1.2$ & $0.73\pm0.15$&$0.26\pm0.02$\\
$L$ ($10^{7} L_\odot$)&$930\pm100$ & $49\pm10$& $2.7\pm0.2$\\
Deconvolved size (FWHM) & $(0^{\prime\prime}.63\pm0^{\prime\prime}.03)\times(0^{\prime\prime}.40\pm0^{\prime\prime}.05) $ & $(0^{\prime\prime}.98\pm0^{\prime\prime}.18)$&---\\ 
Physical size (kpc$^2$) &$(3.7\pm0.2)\times(2.4\pm0.3)$&$5.8\pm1.1$&---\\
$L'$ ($10^{10}$ K km s$^{-1}$ pc$^2$)&$4.3\pm0.5$&$0.49\pm0.10$&$7.0\pm0.5$\\
\hline \noalign {\smallskip}
\end{tabular}
\tablecomments{We produced integrated line maps over the line FWHM and used {\sc casa} to fit a 2D Gaussian model. Then, we extracted aperture spectra utilizing the FWHM of the spatial Gaussian model,  correcting the total flux by a factor of 2 to account for the flux outside the aperture. The deconvolved emission size was measured from Gaussian fitting to the visibilities. $^a$: Fixed to the parameter values derived from the [C{\sc ii}] line. }
\end{table*}

\begin{figure}[tbh]
\centering{
 \includegraphics[width=.5\textwidth]{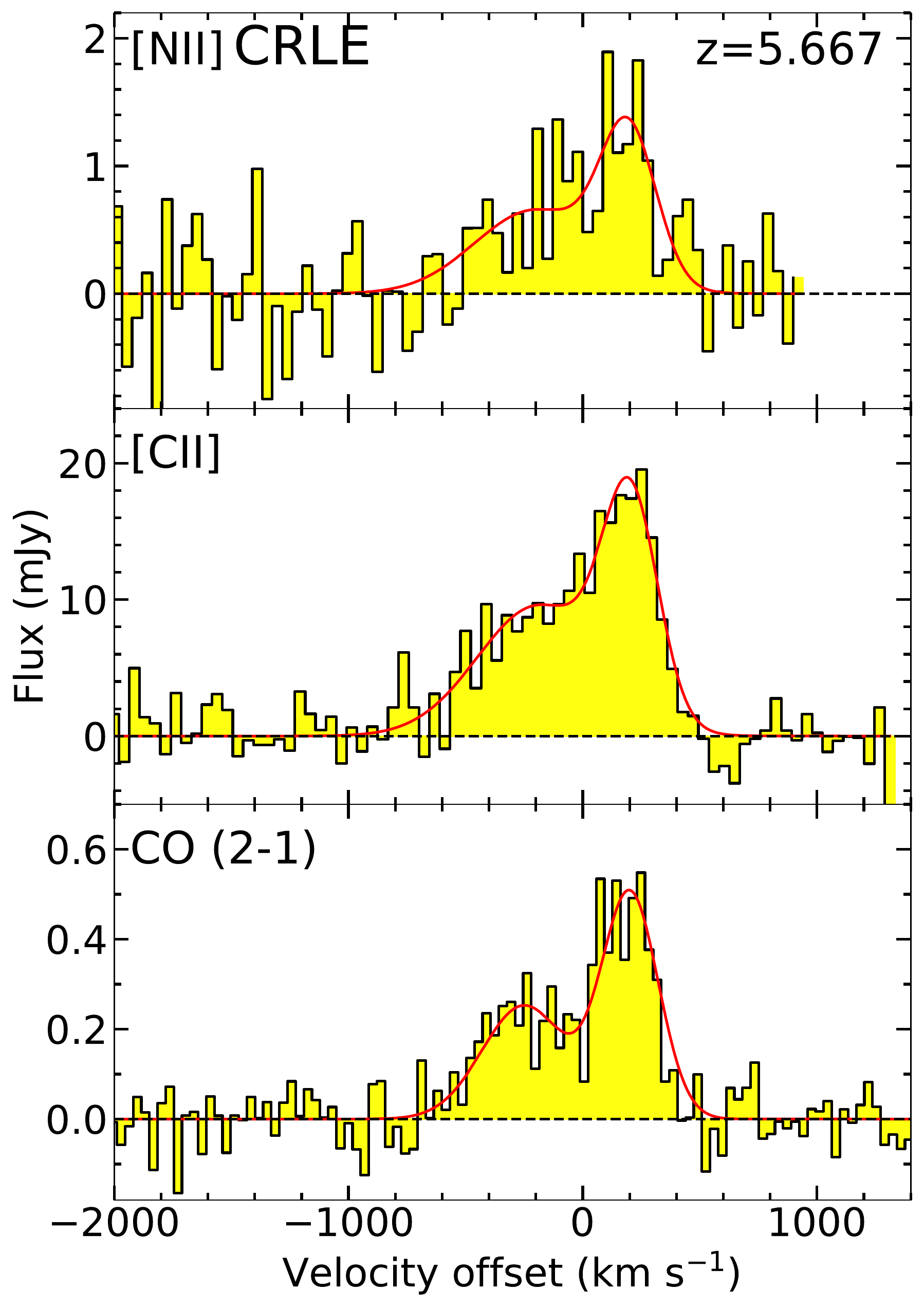}
}
\caption{[N{\sc ii}], [C{\sc ii}], and CO(2--1) continuum-subtracted, aperture integrated line spectra for CRLE (histograms). Double Gaussian fits to the line emission are shown as red curves. A mean redshift of $z=5.667$ was adopted as the velocity reference. The signal-to-noise ratio of the [N{\sc ii}] line does not allow fitting the central frequencies and the widths for the two velocity components. Therefore, these were fixed to the  values measured for [C{\sc ii}]. The channel velocity widths are $\sim$43, $\sim$44 and $\sim$35 km s$^{-1}$ for the [N{\sc ii}], [C{\sc ii}] and CO(2--1) lines, respectively.
}
\label{fig:spectra}
\end{figure}

\section{Line emission properties}
\subsection{Results}
We  detect [C{\sc ii}], [N{\sc ii}] 205$\,\mu$m and CO(2--1) line emission and the  adjacent continuum toward CRLE at  high significance  ($>8\sigma$; Figure~\ref{fig:mom_cont}), which provides an unambiguous redshift identification. Although a foreground galaxy ($z_{\rm phot}\sim0.3$) is located along the line of sight to CRLE, potentially causing gravitational lensing of its emission, in Appendix A we constrain the likely magnitude of this effect to be minor ($<10\%$ flux magnification and negligible spatial distortion).
 Therefore, we simply extract aperture spectra and fit double Gaussians to the spectral profiles to measure the intrinsic line properties (Figure~\ref{fig:spectra}). We report the  line properties in Table~\ref{table_lines}, while the continuum fluxes are listed together with the archival  photometric measurements in Table~\ref{table_HST_flux}.

We measure the deconvolved [C{\sc ii}] spatial FWHM size from  {\sc uv} plane modeling (see Appendix C) to be $0^{\prime\prime}.46\pm0^{\prime\prime}.08$, which corresponds to  $2.7\pm0.5$ kpc at $z=5.667$. Using an isotropic virial estimator \citep{Engel10}, and assuming a [C{\sc ii}] single-Gaussian fit line FWHM of $640\pm60$ km s$^{-1}$ (which is also compatible with the broad velocity component), we derive a dynamical mass of $(1.5\pm0.4)\times10^{11}\,M_\odot$\footnote{We note that this estimate is subject to systematic uncertainties of the order of a factor of a few. In particular, a disk-like gas distribution would require an inclination correction.}.

The [N{\sc ii}]  emission is only slightly resolved.  Using the {\sc casa} task {\sc uvmodelfit}, we measure a deconvolved FWHM major axis size of  $0^{\prime\prime}.98\pm0^{\prime\prime}.18$, corresponding to $5.8\pm1.1$ kpc. 
We  use the same technique to fit the size of the [C{\sc ii}] emission to provide an accurate comparison to the [N{\sc ii}] emission size, obtaining a result that is  compatible with our more sophisticated {\sc uv} plane modeling.
The  [N{\sc ii}] line emission appears to be marginally more extended (formally by a factor of $2.1\pm0.5$) than the [C{\sc ii}] emission, but higher resolution, and higher signal-to-noise [N{\sc ii}] observations are necessary to confirm this finding. In particular, a manual inspection of the {\sc uv}-radial profile of the [N{\sc ii}] line visibilities appears compatible with the size of the [C{\sc ii}] emission.
Neither the CO line nor the adjacent continuum emission are resolved.
We fit the size of the  continuum emission at $158\,\mu$m and $205\,\mu$m in the {\sc uv} plane, and they are compatible with the same deconvolved size of $(0^{\prime\prime}.39\pm0^{\prime\prime}.01)\times(0^{\prime\prime}.31\pm0^{\prime\prime}.02) $.
This implies that the size of the [C{\sc ii}] emitting region is more extended than the size of the dust continuum by $\sim(30\pm20)$\%, in linear dimensions,  suggesting that the observed dust continuum may not represent the full extent of the star-forming gas distribution (see also e.g., \citealt{Riechers14,Chen_resolved}).

We find a [C{\sc ii}]/[N{\sc ii}] line luminosity ratio of $19\pm4$ in CRLE, which is comparable to the line ratio in the $z=5.3$ DSFG AzTEC-3 of $22\pm8$ \citep{Pavesi16}. This line ratio is sensitive to the fraction of [C{\sc ii}] emission coming from ionized gas, rather than photon-dominated regions (PDR), because the ionization energy of nitrogen, in contrast to carbon, is higher than that of hydrogen. This makes it a tracer of ionized gas only, and the [C{\sc ii}]/[N{\sc ii}] ratio is approximately constant in ionized gas (e.g., \citealt{Oberst2006, Pavesi16}).
Assuming a line ratio of $3\pm0.5$ in the ionized gas \citep{Tanio17}, we calculate a fraction of the [C{\sc ii}] emission from PDRs of $84\%\pm4\%$ for CRLE and $86\%\pm5\%$ for AzTEC-3, respectively.

\subsection{Origin of the [C{\sc ii}] emission}
\begin{figure*}[tbh]
\centering{
 \includegraphics[width=\textwidth]{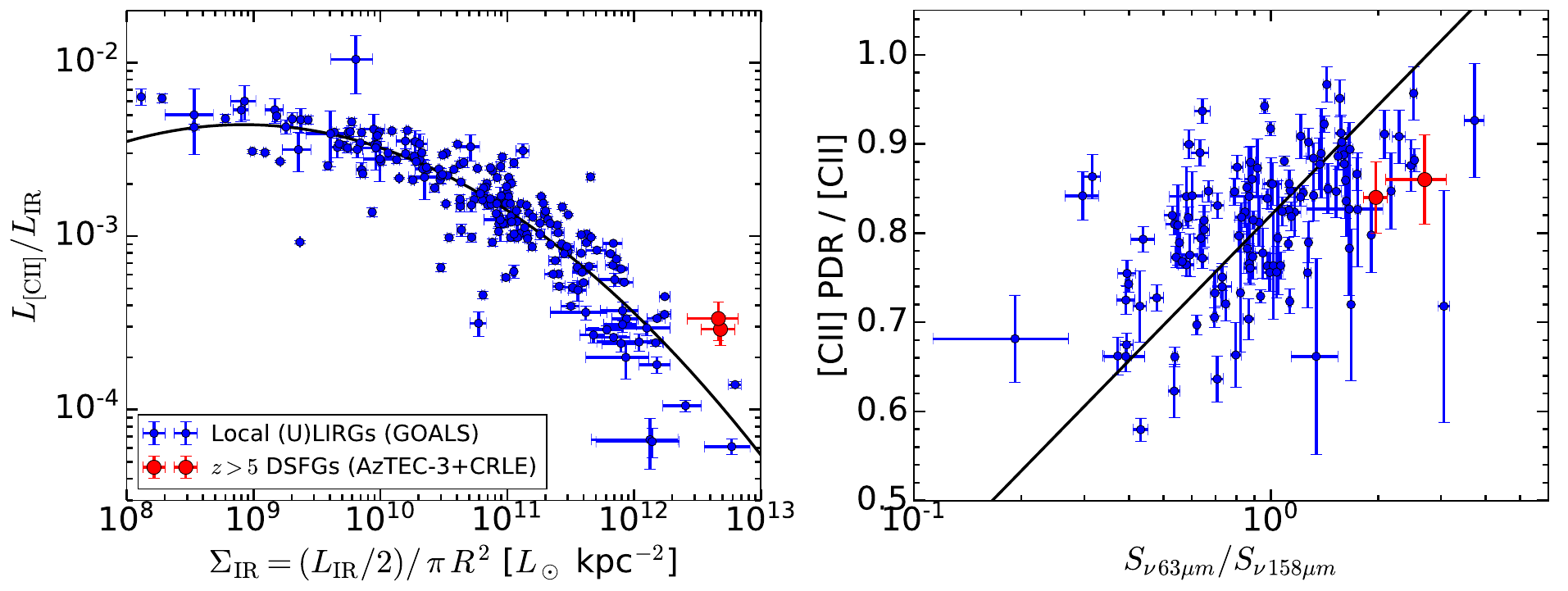}
}
\caption{Left: [C{\sc ii}]-to-IR luminosity ratio as a function of IR luminosity surface density for CRLE and AzTEC-3 (red) and  the GOALS sample of local (U)LIRGs (blue), adapted from \cite{Tanio17}. The black line shows the best fitting function reported by \cite{Tanio17}. Although the $z>5$ DSFGs  show  low [C{\sc ii}] to IR luminosity ratios as expected for extreme starbursts, they appear to show more luminous [C{\sc ii}] emission than might be expected from the local relation. Right: Fraction of the [C{\sc ii}] luminosity coming from PDRs as a function of  FIR color (rest-frame 63 $\mu$m to 158 $\mu$m; a proxy for dust temperature) for CRLE, AzTEC-3 and the GOALS sample \citep{Tanio17}. The black line shows the best fitting function reported by \cite{Tanio17}. The FIR colors for the $z>5$ DSFGs were inferred from the best-fitting modified-blackbody models. The PDR fraction of the [C{\sc ii}] emission was estimated from the [C{\sc ii}]/[N{\sc ii}] line ratio, assuming a line ratio of $3\pm0.5$ in the ionized gas, following \cite{Tanio17}. The $z>5$ DSFGs appear to show a lower PDR fraction of the [C{\sc ii}] emission relative to the local trend with dust temperature.
}
\label{fig:tanio}
\end{figure*}

 CRLE exhibits approximately the same [C{\sc ii}]/[N{\sc ii}] ratio as AzTEC-3 \citep{Pavesi16}. This ratio is compatible with the range observed in local (ultra)luminous infrared galaxies ((U)LIRGs). However, given the higher gas density and star formation rate surface density in high redshift DSFGs relative to most local (U)LIRGs, one might expect a lower fraction of the [C{\sc ii}] emission to come from the diffuse ionized gas, which is traced by [N{\sc ii}], relative to atomic and molecular PDRs and therefore a higher [C{\sc ii}]/[N{\sc ii}] ratio \citep{Pavesi16}. Therefore, our observed line ratio suggests that high redshift DSFGs may host a significant reservoir of diffuse ionized gas, which may be more extended than the starbursting gas. The potentially larger size of the [N{\sc ii}] emitting region in CRLE, relative to  [C{\sc ii}]  (although with significant uncertainty), would provide support for this interpretation if confirmed. Recent measurements of both [N{\sc ii}] fine-structure lines at 205$\,\mu$m (studied here), and 122$\,\mu$m in local galaxies also support the interpretation of the [N{\sc ii}] emission predominantly coming from diffuse ionized gas, rather than compact H{\sc ii} regions (e.g., \citealt{HerCam16,Shinings1,Shinings2,Tanio17}).

 A well known decreasing trend of the [C{\sc ii}]/$L_{\rm IR}$ ratio with infrared luminosity surface density (a proxy for FUV field strength) offers insight into the origin of the [C{\sc ii}] emission. In particular, a decreasing trend is expected for PDR emission due to the saturation of [C{\sc ii}] line emission at higher FUV field strengths (e.g., \citealt{Stacey91}).
This trend is clearly observed in a sample of local dusty star-forming galaxies from the Great Observatories All-sky LIRG Survey (GOALS; \citealt{Tanio17}). However, at the calculated infrared luminosity surface density for CRLE and AzTEC-3, their observed [C{\sc ii}]/$L_{\rm IR}$ ratios ($\sim3-4\times10^{-4}$) lie significantly above the best-fit to the relation for the GOALS sample (Figure~\ref{fig:tanio}).

This may suggest an additional contribution besides PDRs to the [C{\sc ii}] emission.
 If we assume that the PDR contribution to the [C{\sc ii}] luminosity is uniquely determined by the intensity of the radiation field as traced by the $L_{\rm IR}$ surface density, a more significant contribution from diffuse ionized gas may be a reason for this excess.
Furthermore, the fraction of [C{\sc ii}] coming from PDR, rather than ionized gas (as traced by the [C{\sc ii}]/[N{\sc ii}] ratio), appears to be lower for CRLE and AzTEC-3 for their modeled FIR color, relative to the fitted relation for the GOALS sample (Figure~\ref{fig:tanio}). This quantifies the finding that the [C{\sc ii}]/[N{\sc ii}] ratio appears to be low, for the more extreme (i.e., higher dust temperature, Figure~\ref{fig:tanio}) gas conditions observed in these $z>5$ DSFGs relative to local starbursts \citep{Pavesi16}. As such, it appears to point towards a larger contribution from diffuse ionized gas to the [C{\sc ii}] emission than what is expected from local starbursts. This diffuse ionized gas may take the form of a gas halo, more extended than the star-forming region. We suggest that a potentially larger and more diffuse ionized gas component at $z>5$  may be due to freshly accreted, inflowing gas, which is expected to play an important role in powering starbursts in the first billion years of cosmic time.
Metallicity is unlikely to play a significant role in affecting the [C{\sc ii}]/[N{\sc ii}] line ratio, as we constrain $\alpha_{\rm CO}$ to be low, and the high dust mass for CRLE (see below) likely implies a solar or super-solar metallicity \citep{Bolatto_review}. Nonetheless, an important caveat to our conclusions is that variations in the carbon to nitrogen abundance ratio may affect our interpretation of the [N{\sc ii}] line emission properties.

\section{Spectral Energy Distribution Analysis}

\subsection{Optical-to-NIR SED }

\begin{table}[t]
\centering
\caption[]{Measured continuum fluxes (foreground galaxy \& CRLE).}
\label{table_HST_flux}
\begin{tabular}{ c | c | c}
\hline 
  Wavelength ($\mu$m) & Flux ($\mu$Jy)& Band \\

\hline \noalign{}
0.4816&$4.73	\pm0.03$ & {\it HSC} g\\				
0.6234&$10.49	\pm0.03$& {\it HSC} r\\
0.7740&$13.38\pm	0.04$& {\it HSC} i\\
0.9125&$17.44	\pm0.06$& {\it HSC} z\\
0.9780&$17.15\pm	0.12$& {\it HSC} Y\\
\hline \noalign{}
1.0552  & $15.15 \pm 0.07$ &{\it HST}/F105W \\
1.2501   & $17.20 \pm 0.06$&{\it HST}/F125W\\
1.5418   &  $20.67 \pm  0.08$ &{\it HST}/F160W\\
\hline \noalign{}
3.6 &$16.29\pm0.17$& {\it Spitzer}/IRAC\\
4.5 &$15.86\pm0.18$& ---\\
5.8  &$19\pm5$& ---\\
 8.0 &$21\pm4$ &---\\
24 & $70\pm15$ & {\it Spitzer}/MIPS\\
\hline \noalign {}
 100$^*$&$<5,000$ &{\it Herschel}/PACS \\
 160$^*$ &$<10,000$&---\\
 250$^*$&$12,000\pm900$&{\it Herschel}/SPIRE\\
 350$^*$&$20,900\pm1,300$&---\\
 500$^*$&$31,100 \pm1,400$&---\\
\hline \noalign {}
850$^*$&$16,700\pm2,000$& SCUBA-2\\
\hline \noalign {}
1,024 (292.8 GHz)$^*$ & $16,500\pm900$ &ALMA\\
1,308 (229.2 GHz)$^*$& $8,650\pm300$ &---\\
8,800 (34.069 GHz)$^*$ & $22\pm2$&VLA\\
\hline \noalign {}

\end{tabular}

 \begin{tablenotes}
 \small
 \item 
\textbf{Note} The {\em HST}/WFC3 fluxes measured with SExtractor, using the Kron mode. We report the {\it Herschel}/PACS non-detection as $3\sigma$ upper limits \citep{PEP}. The SCUBA-2 flux was measured from the S2CLS images \citep{Geach17} and the other fluxes were obtained from the COSMOS2015 catalog \citep{Laigle16}. $^*$: Measurements expected to be dominated by emission from CRLE. Additional  fluxes may be found in the publicly available COSMOS2015 catalog.
 \end{tablenotes}
\end{table}

\begin{figure*}[tbh]
\centering{

\includegraphics[width=.7\textwidth] {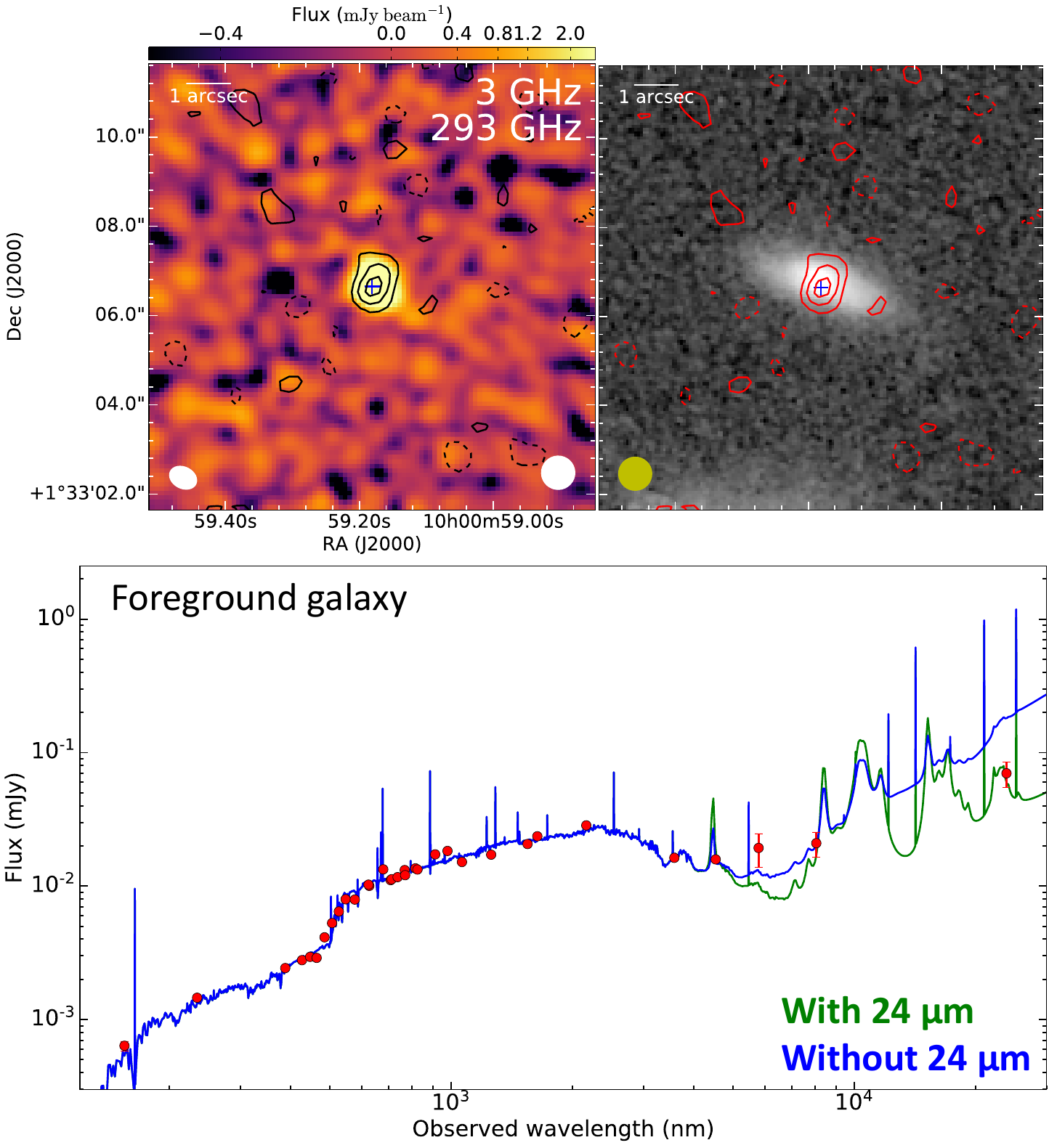}
}
\caption{Top: 3 GHz observed-frame continuum emission from CRLE (contours; \citealt{Smolcic17}) on top of ALMA dust continuum from CRLE (left), and NIR emission from the foreground galaxy (right). Contours are shown in $2\sigma$ steps, starting from $\pm2\sigma$.  The blue cross represents the $158\,\mu$m dust  emission peak. Top left:  Color-scale, showing the rest-frame 158$\,\mu$m continuum (corresponding to observed-frame 293 GHz). The  beam sizes shown are for the 293 GHz data (left), and the 3 GHz data (right).   Top right: Gray-scale {\em HST}/WFC3 F160W image from \cite{Barisic17}, showing the foreground disk galaxy at a photometric redshift $z\sim0.35$. The contribution from CRLE cannot  reliably be separated from the foreground galaxy at these wavelengths.
Bottom: UV-NIR SED of the foreground galaxy and CRLE, fitted with {\it Cigale} models.  The  models do not provide a good fit to the   measurement at $5.7\,\mu$m, suggesting that emission from CRLE may contribute a non-negligible fraction of, at least, the $5.7\,\mu$m flux.
}
\label{fig:foreg_SED}
\end{figure*}

 CRLE is located behind a local foreground spiral galaxy with a  photometric redshift of $\sim0.35$ and stellar mass of $3.3\times10^9\,M_\odot$ in the COSMOS2015 catalog \citep{Laigle16}, which outshines its optical emission. 
We analyze the {\em HST}/WFC3 data targeting HZ10  \citep{Barisic17} and measure the  NIR fluxes for the foreground galaxy in the Y, J and H bands, potentially including a contribution from CRLE (Figure~\ref{fig:foreg_SED}).
We measure the fluxes in the {\em HST} images, adopting Kron fluxes as measured by {\it Source Extractor} (Table~\ref{table_HST_flux}; \citealt{Sextractor}). We then fit the foreground galaxy with {\it Galfit} \citep{Galfit} and find a faint residual flux to the west of its center. We fix the model to be an exponential light profile with axis ratio of 0.5 and fit for center position, size and flux.
The residual flux in the H band data corresponds to only $\sim1-4$\% of the total emission, which may be associated with CRLE, or with an imperfect fit to the central regions of the foreground galaxy (see Appendix B for further details).
The foreground galaxy is therefore expected to dominate the total emission at optical-to-NIR wavelengths, at least up to observed-frame $\sim2\,\mu$m. 
In order to constrain the contribution  of CRLE to the total emission in the near to mid-IR, we use {\it Cigale} to carry out spectral energy distribution (SED) fitting of the optical and NIR emission \citep{Cigale1,Cigale2}. 
We fit the optical and NIR stellar emission with a delayed star formation history (with ages of the oldest stars ranging from 250 Myr to 12 Gyr, and an e-folding time ranging from 50 Myr to 8 Gyr), stellar population synthesis models with metallicity ranging from $1/50\,Z_\odot$ to $Z_\odot$ according to \cite{BruzualCharlot} and a dust attenuation law according to \cite{Calzetti}.
We determine a stellar mass of  $(3.0\pm0.5)\times10^9\,M_\odot$ and a star formation rate of $\sim1.5\,M_\odot\,$ yr$^{-1}$ for the foreground galaxy, confirming the catalog stellar mass value.  We use the best fitting {\it Cigale} models to constrain the FIR luminosity of the foreground galaxy by adopting dust emission models according to \cite{DL07}. We find approximate predictions for the {\it Herschel}/SPIRE fluxes of 3.7, 2.0, and 0.8 mJy, in the 250, 350 and 500 $\mu$m bands, respectively. These are significantly lower than the measured fluxes in these bands.  Therefore, we consider the FIR emission from the foreground galaxy to be subdominant to the emission from CRLE (Table~\ref{table_HST_flux}).

 The {\it Spitzer}/IRAC 5.8 and 8.0$\,\mu$m and MIPS $24\,\mu$m fluxes are not  fit well by the available SED models (Figure~\ref{fig:foreg_SED}).  We interpret this to be an indication of significant contamination by  emission from CRLE toward the longer wavelengths.
Although the IRAC1 and IRAC2 fluxes are compatible with expectations, the higher IRAC3 and IRAC4 fluxes are difficult to reproduce with the adopted  dust emission models \citep{DL07}. The  best-fitting SED models were obtained by either including or excluding the MIPS $24\,\mu$m flux in the fitting (Figure~\ref{fig:foreg_SED}). We find that best-fitting models to the IRAC points over-predict the $24\mu$m flux when excluded, and that best-fitting models that include the $24\mu$m measurement require strong PAH emission to fit the IRAC4 flux and they under-predict the IRAC3 flux at 5.8$\,\mu$m by at least a factor $\sim2$. Even artificially reducing the measurement uncertainty on the IRAC3 flux does not improve the fit to this measurement.
While it is unclear if the $24\mu$m flux is dominated by emission from CRLE or from the foreground galaxy, we interpret these results as suggesting that perhaps a significant fraction of the flux at 5.8$\,\mu$m may be due to CRLE. A potential caveat to this conclusion may arise in case the foreground galaxy contains an  AGN, which could contribute additional flux at these wavelengths, and is presently unconstrained.
We use {\it Cigale} to derive approximate constraints to  the stellar mass in CRLE. If all the IRAC3 and IRAC4 emission were due to CRLE,  the implied stellar mass would be of order $\sim1-2\times10^{11}\,M_\odot$. This likely provides an upper limit to the actual stellar mass of CRLE. A contribution of $\sim$25\%--50\% of the emission may be plausible, yielding a best stellar mass estimate in the range of $\sim2.5-10\times10^{10}\,M_\odot$.



\subsection{FIR SED and modified blackbody fitting}

\begin{figure}[htb]
\centering{
 \includegraphics[width=.45\textwidth]{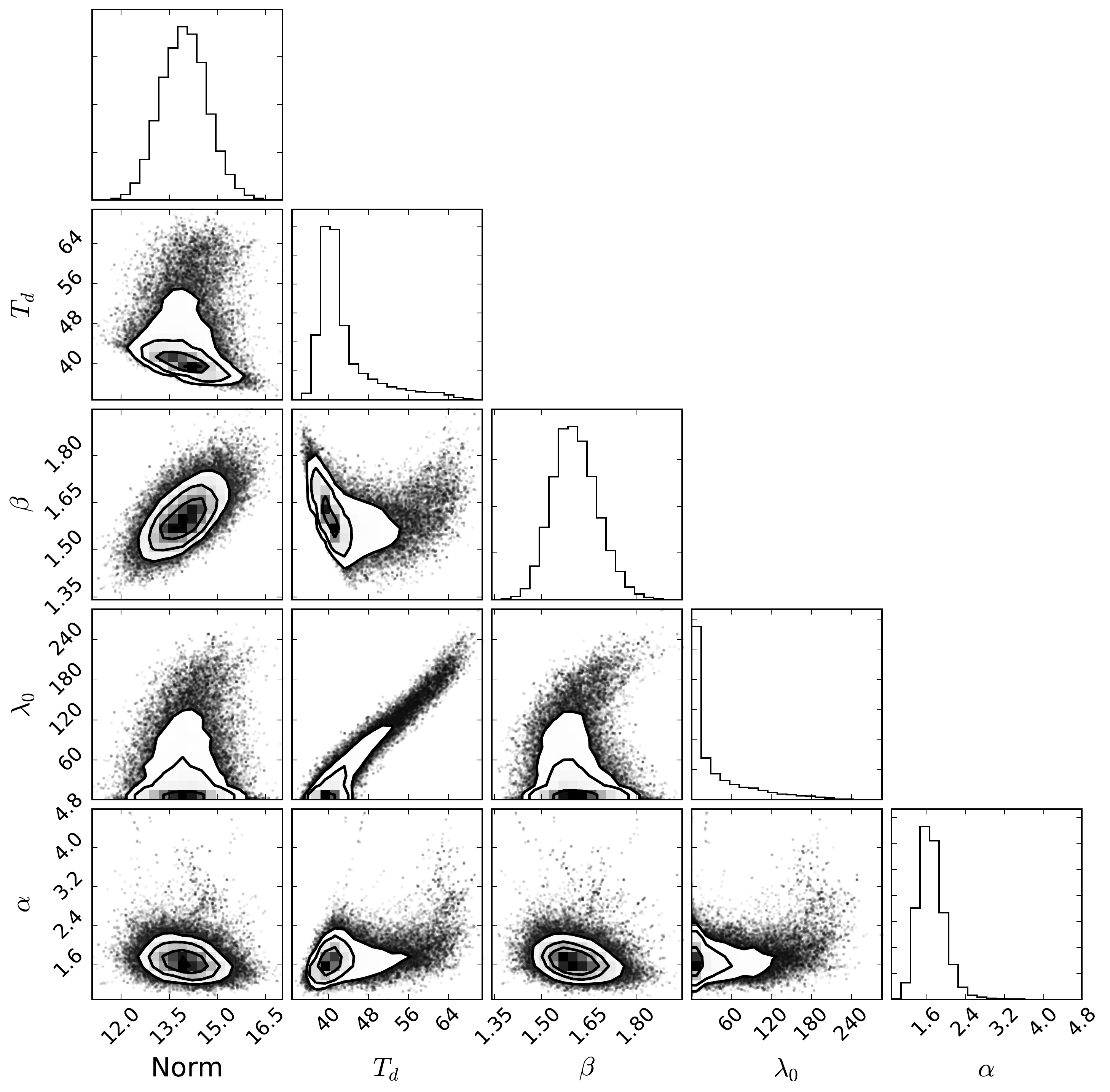}
  \includegraphics[width=.5\textwidth]{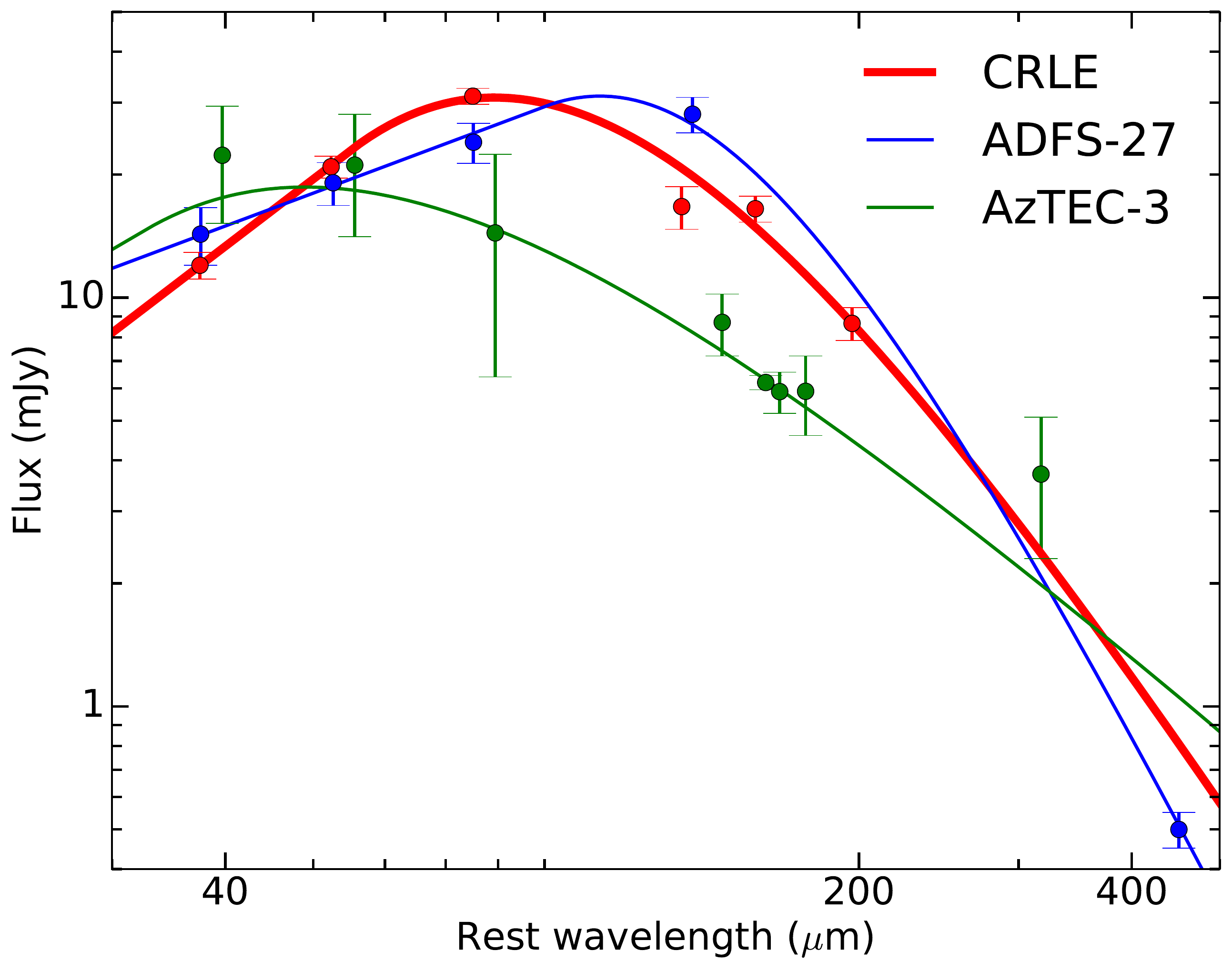}
  \includegraphics[width=.5\textwidth]{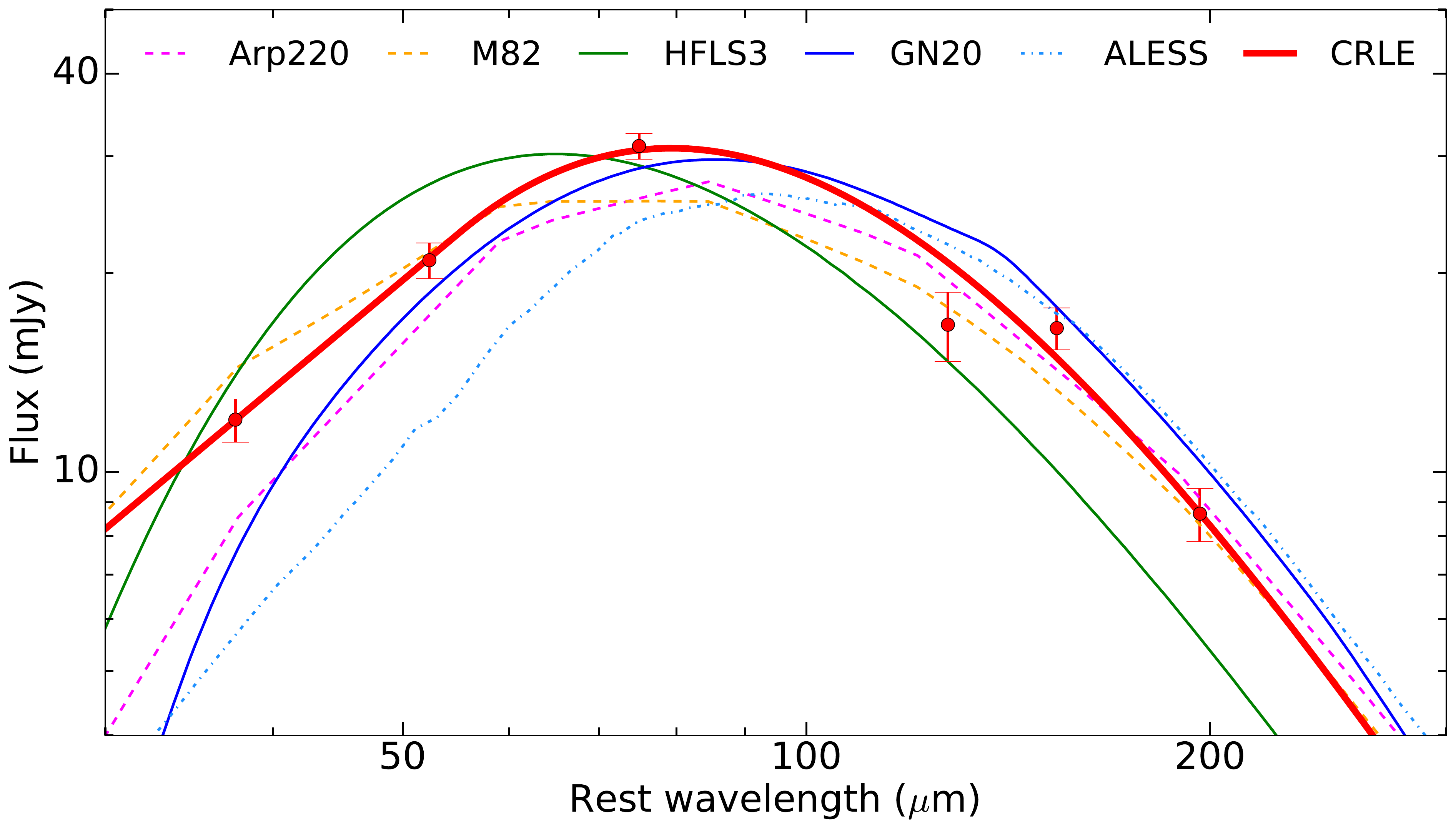}
}
\caption{Modified blackbody (MBB) fits to the FIR SED in CRLE. Top: Corner plot of the model parameter posterior distribution.  Middle: FIR SED comparison to other $z>5$ DSFGs (AzTEC-3 and ADFS-27; \citealt{Riechers14,Riechers17}). CRLE shows peak emission at intermediate rest-wavelengths between the hot, compact starburst AzTEC-3 and the early-stage merger ADFS-27. Bottom:  Comparison to other low and high redshift FIR SED templates \citep{Silva98,RiechersHFLS3,Magdis11,Swinbank14}.}
\label{fig:MBB_fits}
\end{figure}

In order to model the FIR SED of CRLE as measured by {\it Herschel}, ALMA, and JCMT/SCUBA-2, we use a modified blackbody, smoothly connected to a mid-IR power-law. We use {\em emcee} \citep{emcee} to explore the posterior probability distribution of the parameters shown in Fig.~\ref{fig:MBB_fits}. We employ uniform, non-constraining priors for the parameters (i.e., flux normalization at rest-frame 158 $\mu$m,  dust temperature $T_d$,  dust emissivity  parameter $\beta$,  rest-frame wavelength at which the optical depth becomes unity $\lambda_0$, and  mid-IR power law index $\alpha$; Table~\ref{table_MBB}).
By integrating between 42.5 and 122.5$\,\mu$m, we find a FIR luminosity of $L_{\rm FIR}=(1.55\pm0.05)\times10^{13}\,L_\odot$. By integrating between 8 and 1000$\,\mu$m, we derive an IR luminosity of $L_{\rm IR}=(3.2\pm0.3)\times10^{13}\,L_\odot$. 
Because  the SED of CRLE is not constrained at mid-IR wavelengths, we follow the standard practice of estimating the star formation rate based on the FIR luminosity only, to provide a comparison to other $z>5$ DSFGs \citep{Riechers14}. By adopting the standard conversion based on a Chabrier IMF \citep{CarilliWalter}, we infer a star formation rate of $(1550\pm50)\,M_\odot$ yr$^{-1}$, with the caveat that the real star formation rate may be up to a factor of $\sim2\times$ higher.  Given the high level of dust obscuration in CRLE, we cannot exclude the presence of an obscured AGN. However, we expect that an AGN would only introduce minor contributions to the rest-frame $>42.5\,\mu$m luminosity given the high dust-obscured star formation rate of CRLE (see, e.g., \citealt{Kirkpatrick15}). As an example, the dust-obscured AGN in the $z=4.05$ DSFG GN20 may dominate the mid-IR luminosity, but its contribution to the total IR luminosity appears to be minor \citep{Riechers_GN20}.
Even for luminous high-redshift quasars, the contribution from AGN-heated, hot dust to the far-IR luminosity does not appear to exceed $\sim20\%$ \citep{Leipski13,Leipski14}.
By adopting standard (although uncertain) assumptions from \cite{Dunne03}, we find an estimated dust mass of $(1.3\pm0.3)\times10^9\,M_\odot$.

The dust SED shape for CRLE appears to be intermediate between that of AzTEC-3 ($z\sim5.3$) and ADFS-27 ($z\sim5.7$; Figure~\ref{fig:MBB_fits}, \citealt{Riechers14, Riechers17}). We also compare the measured FIR SED of CRLE to model templates of selected low- and high-redshift starbursts, spanning a range of physical conditions.
A comparison to the template for local starbursts Arp 220 and M82 shows that CRLE resembles M82 more closely, suggesting a comparatively low dust optical depth\footnote{The high optical depth in Arp 220 characteristically suppresses the emission at the short wavelengths, reducing the observable emission from the hot dust component  (e.g., \citealt{NickArp220}).} \citep{Silva98}. A comparison to the ultra luminous $z=6.34$ DSFG HFLS3 shows that CRLE appears to have a significantly lower dust temperature \citep{RiechersHFLS3, Ivison16}, as evidenced by the longer wavelength of the peak emission, while at the same time displaying a warmer dust temperature than is observed in the lower redshift ALESS sources \citep{Swinbank14}. The SED of CRLE appears to closely resemble that of GN20, an extended $z\sim4$ DSFG (Figure~\ref{fig:MBB_fits}, e.g.,  \citealt{Carilli10,Magdis11,Hodge12,Hodge15}).  Our reference SED templates suggest a  potential redshift trend toward hotter dust in higher redshift DSFGs, as evidenced by the shorter wavelength  SED peak in CRLE relative to the ALESS sample average (mostly in the range $z\sim2-3$, \citealt{Danielson17}),  and the longer wavelength in comparison to HFLS3 \mbox{($z$=6.34, \citealt{RiechersHFLS3, Faisst17})}. Heating from a warmer cosmic microwave background (CMB) may partly contribute to this tentative dust temperature trend (e.g., \citealt{daCunhaCMB}).
However, selection effects may also be partly responsible for this trend, since most of these DSFGs were selected at fixed observed-frame wavelengths.

\begin{table}[t]
\caption[]{Results from modified blackbody fitting to the FIR SED of CRLE.}
\label{table_MBB}
\resizebox{1\columnwidth}{!}{%
\begin{tabular}{ c c c c c c }
\hline 
Percentile & Norm. (158$\,\mu$m)& $T_d$ & $\beta$&rest $\lambda_0$&$\alpha$\\
 & (mJy)&(K)&&($\mu$m)&\\
\hline \noalign {\smallskip}
16th&13.25& 38.95&   1.537&  1.659&  1.430\\
50th&13.94&  41.16&   1.604&  15.77&   1.658\\
84th& 14.65 &  47.49&   1.674&  80.08&   1.938\\
\hline \noalign {\smallskip}
\end{tabular}
}
\end{table}

\subsection{Radio  continuum emission}

CRLE is detected at 7$\sigma$ significance in  3 GHz continuum emission ($24.3\pm3.8\,\mu$Jy; \citealt{Smolcic17}).
The emission is not resolved at a synthesized beam size of $0.75^{\prime\prime}$ and it is aligned with the  dust continuum emission (Fig.~\ref{fig:foreg_SED}).
CRLE is also potentially weakly detected at 1.4 GHz at the $2.5\sigma$ level \citep{Schinnerer10}.
 We conservatively adopt a $3\sigma$ limit of  $<84\,\mu$Jy at 1.4 GHz, which is consistent with a spectral index of $-0.7$.
Sensitive low radio frequency observations also provide constraining non detections (Tisani\'c  et al., in prep.). We derive $3\sigma$ upper limits of 320 $\mu$Jy at 325 MHz and 150 $\mu$Jy at 610 MHz. These imply constraints to the effective radio spectral index to observed-frame 3 GHz of $>-1.16$ and $>-1.14$ for the 325 MHz and 610 MHz limits, respectively. 
 

Adopting the redshift-dependent FIR-radio correlation measured by \cite{Delhaize17}, we can convert our measured 3 GHz flux and the constraints to the 1.4 GHz flux to FIR luminosities. The FIR-radio correlation has been calibrated at a rest-frame frequency of 1.4 GHz, which corresponds to observed-frame 210 MHz for CRLE. This comparison therefore requires significant extrapolation from our measurement at 3 GHz, and it is sensitive to the spectral index.
 Adopting a radio power law index of $-0.7$ as in \cite{Delhaize17}, we use  3 GHz flux and  1.4 GHz flux upper limit  to derive FIR luminosities of $4.0\times10^{12}\,L_\odot$ and  $<8.0\times10^{12}\,L_\odot$, respectively.  
This radio-inferred FIR luminosity is significantly lower than our direct measurement. We could reconcile these estimates with an effective spectral index of $-1.2$ between rest-frame 20 GHz and 1.4 GHz, but this would be in slight tension with our 325 and 610 MHz upper limits.
Alternatively, \cite{Molnar} have suggested that the FIR-radio correlation may not evolve in star-forming, disk-dominated galaxies. If we assume no redshift evolution, the 3 GHz flux and the 1.4 GHz flux upper limit would  imply FIR luminosities of $\sim2.7\times10^{13}\,L_\odot$  and $<5.5\times10^{13}\,L_\odot$, in agreement with our direct measurement.

However, the observed radio emission may  not be well described by a single power-law spectral index down to rest-frame 1.4 GHz. In particular, the analysis by \cite{Tabatabaei17} suggests that approximately half of the radio flux at this frequency may be due to thermal free-free emission. Under this assumption, and adopting the relationship between thermal radio emission and star-formation rate  by \cite{Murphy11} for an H{\sc ii} region electron temperature of $T_e=10^4\,K$, we would infer a SFR$\sim4000\,M_\odot\,$yr$^{-1}$ from the 3 GHz continuum flux, with large uncertainties due to assuming a thermal versus non-thermal fraction.

\subsection{Star-forming gas properties}
Low-{\it J} CO line luminosities are expected to provide a reliable estimate of the molecular gas mass (e.g., \citealt{Bolatto_review}). Here, we  assume a brightness temperature ratio of $R_{21}=1$ between the CO {\em J}=2--1 and 1--0 lines, due to the moderately high inferred dust temperature of CRLE (Table~\ref{table_MBB}).\footnote{Previous samples of DSFGs at lower redshift show nearly thermalized gas excitation up to the {\em J}=2--1 transition, justifying our assumption ($R_{21}\sim0.85-0.95$; e.g., \citealt{CarilliWalter}).}
Another effect which may be relevant to the interpretation of the observed CO line flux is contrast against the warmer CMB, and the additional gas heating this provides at $z>5$ \citep{daCunhaCMB}. While the effect this may introduce is difficult to estimate without additional CO excitation constraints, \cite{daCunhaCMB} suggest that for warm gas kinetic temperatures, and moderately high gas densities we may expect the observed CO line flux to be suppressed by $\sim$0.5--0.8, at this redshift. We do not attempt to estimate a  correction factor for this effect, but include it  into the definition of $\alpha_{\rm CO}$ instead (i.e., a larger CMB correction would imply a higher $\alpha_{\rm CO}$). 
By assuming the dynamical mass to be completely accounted for by molecular gas, we can derive a conservative upper limit on the CO conversion factor of $\alpha_{\rm CO}<2.1\pm0.6\,M_\odot\,({\rm K\, km\,s^{-1}\, pc}^2)^{-1}$. This conversion factor is conservative, because it does not include the stellar contribution to the dynamical mass. Our stellar mass estimate is too uncertain to provide a useful constraint in this context.
Such low conversion factors are typically only observed in highly metal-enriched galaxies, which appears to suggest that metallicity is not a major source of uncertainty in the interpretation of line ratios \citep{Bolatto_review}. Furthermore, from the dust mass estimates presented in Section~4.2, we can adopt a gas-to-dust ratio in order to derive an independent estimate of  $\alpha_{\rm CO}$. The gas-to-dust ratio may be assumed to follow the same dependence on metallicity as measured at lower redshift. Here, we follow \cite{Magdis11} by adopting a super-solar metallicity, expressed as 12+$\log$(O/H), between 8.8 and 9.2. This implies a gas-to-dust mass ratio of 35--75, which is consistent with an average value for DSFGs of $\sim50$ \citep{Santini10}. This range implies a range of $\alpha_{\rm CO}$ from 0.65$\pm0.16 \,M_\odot\,({\rm K\, km\,s^{-1}\, pc}^2)^{-1}$ to 1.4$\pm0.3 \,M_\odot\,({\rm K\, km\,s^{-1}\, pc}^2)^{-1}$. These ranges are consistent with expectations from local ULIRGs, for which a conversion factor of $\alpha_{\rm CO}\sim0.8\,M_\odot\,({\rm K\, km\,s^{-1}\, pc}^2)^{-1}$ was estimated \citep{DownesSolomon}. In the following we  adopt a conversion factor of $\alpha_{\rm CO}=1\,M_\odot\,({\rm K\, km\,s^{-1}\, pc}^2)^{-1}$, which provides a conservative estimate of the molecular gas mass in CRLE\footnote{We note that there remain substantial systematic uncertainties regarding the appropriate CO conversion factor for high redshift DSFGs, as well as for local ULIRGs (e.g., \citealt{Scoville16,NickArp220}). Previous assumptions of $\alpha_{\rm CO}$ may also affect the adopted gas-to-dust ratios, and may therefore be partly responsible for the apparent consistency of our estimates.}. We thus derive a total molecular gas mass of $M_{\rm gas}= (7.0\pm0.5)\times10^{10} \,M_\odot$,  corresponding to $\sim50$\%$\pm15$\% of the dynamical mass estimate.   This gas mass estimate is similar to the gas masses of other $z>5$ DSFGs such as AzTEC-3 ($\sim$5.3$\times10^{10}\,M_\odot$) and HFLS3 ($\sim$4.5$\times10^{10}\,M_\odot$; \citealt{Riechers10a,RiechersHFLS3,Cooray14}).  Our inferred $\alpha_{\rm CO}$ is compatible with model predictions by \cite{Vallini18} at $z\sim6$, which suggest $\alpha_{\rm CO}=(1.5\pm0.9)\,M_\odot\,({\rm K\, km\,s^{-1}\, pc}^2)^{-1}$, despite the sub-solar metallicity of their simulated galaxy. They interpret this low conversion factor as the result of the high density and high turbulence in the molecular gas of their simulated galaxy. The measured [C{\sc ii}]/CO(2--1)  line ratio in CRLE is also compatible with their model predictions, although we caution that their simulated galaxy targets the ``normal" star-forming population with a star formation rate over an order of magnitude lower than CRLE \citep{Vallini18}.

We also use single-wavelength, dust continuum  emission measurements on the Rayleigh-Jeans tail to explore alternative methods to estimate the ISM mass of CRLE. Using the  the rest-frame 205$\,\mu$m  and  1.3 mm observations, we follow \cite{Scoville16,Scoville17} to estimate total ISM masses of  $1.1\times10^{12}\,M_\odot$ and $8.0\times10^{11}\,M_\odot$, respectively. This method was calibrated at a rest-frame wavelength of 850 $\mu$m, hence the latter estimate is expected to be more reliable, since the extrapolation is smaller. We also use rest-frame 500 and 250$\,\mu$m fluxes in CRLE to infer a gas mass following \cite{Groves15}. The rest-frame 500 and 250$\,\mu$m fluxes are estimated from our best-fitting modified black body model, and imply gas masses of $1.3\times10^{12}\,M_\odot$ and $2.4\times10^{11}\,M_\odot$, respectively.  Therefore, these dust-based methods would imply very large ISM masses, exceeding our dynamical mass estimate and suggesting that hotter dust temperatures may affect these techniques making them less reliable for this type of high-redshift galaxy (e.g., \citealt{Scoville15}).

Adopting the relationship between CO luminosity and star formation rate (i.e., the star formation law) for local starburst galaxies \citep{Kennicutt98,CarilliWalter},  we can convert the measured $L'_{\rm CO}$ to an IR luminosity of $L_{\rm IR}=1.3\times10^{13}\,L_\odot$ with large uncertainties  (the scatter of this relationship is a factor $\sim2.5$). This is consistent with our measured IR luminosity, although apparently at the low end of the confidence interval. 
By adopting the fiducial values for the star formation rate and the molecular gas mass we derive a gas depletion time scale of $45\pm4\,$ Myr, which is comparable to  estimates for other $z>5$ DSFGs such as AzTEC-3 ($\sim$50 Myr) and HFLS3 ($\sim$36 Myr) \citep{Riechers10a,RiechersHFLS3,Riechers14}.

\subsection{Merger stage of CRLE}

As shown in Section~4.2, the dust SED in CRLE appears to be intermediate between those of other $z>5$ DSFGs such as AzTEC-3 and ADFS-27 (Fig.~\ref{fig:MBB_fits}; \citealt{Riechers10a,Riechers14, Riechers17}). A possible interpretation of these differences in the dust properties (e.g., temperature or optical depth) among these galaxies involves the merger stage. In particular, AzTEC-3 appears to be a dispersion dominated starburst with an exceptionally compact central density profile which is suggestive of a late stage merger, after coalescence (Riechers et al., in prep.). On the other hand, ADFS-27 is an interacting galaxy pair, where the individual galaxies are separated by just 9 kpc, and therefore may represent an early merger stage \citep{Riechers17}. In Appendix C, we carry out dynamical modeling of the [C{\sc ii}] emission, which suggests that a simple rotating disk model does not provide a good fit to CRLE. This conclusion is also supported by the asymmetric profile visible in the line spectra (Fig.~\ref{fig:spectra}). We find evidence for a second dynamical component, which may suggest an overall dispersion dominated system. This evidence appears to suggest  an ongoing merger in CRLE, with a component separation  of $2.0\pm0.4$ kpc (from the [C{\sc ii}] emission modeling).
Therefore, we tentatively interpret the intermediate SED in CRLE as a reflection of its merger stage.
However, the SED in CRLE also resembles that in GN20, which shows a coherently rotating massive gas disk and shows no evidence for a galaxy merger \citep{Hodge12}.

\section{Galaxy Overdensity around CRLE}

\subsection{Overdensity characterization}

\begin{figure*}[htb]
\centering{
\includegraphics[width=\textwidth]{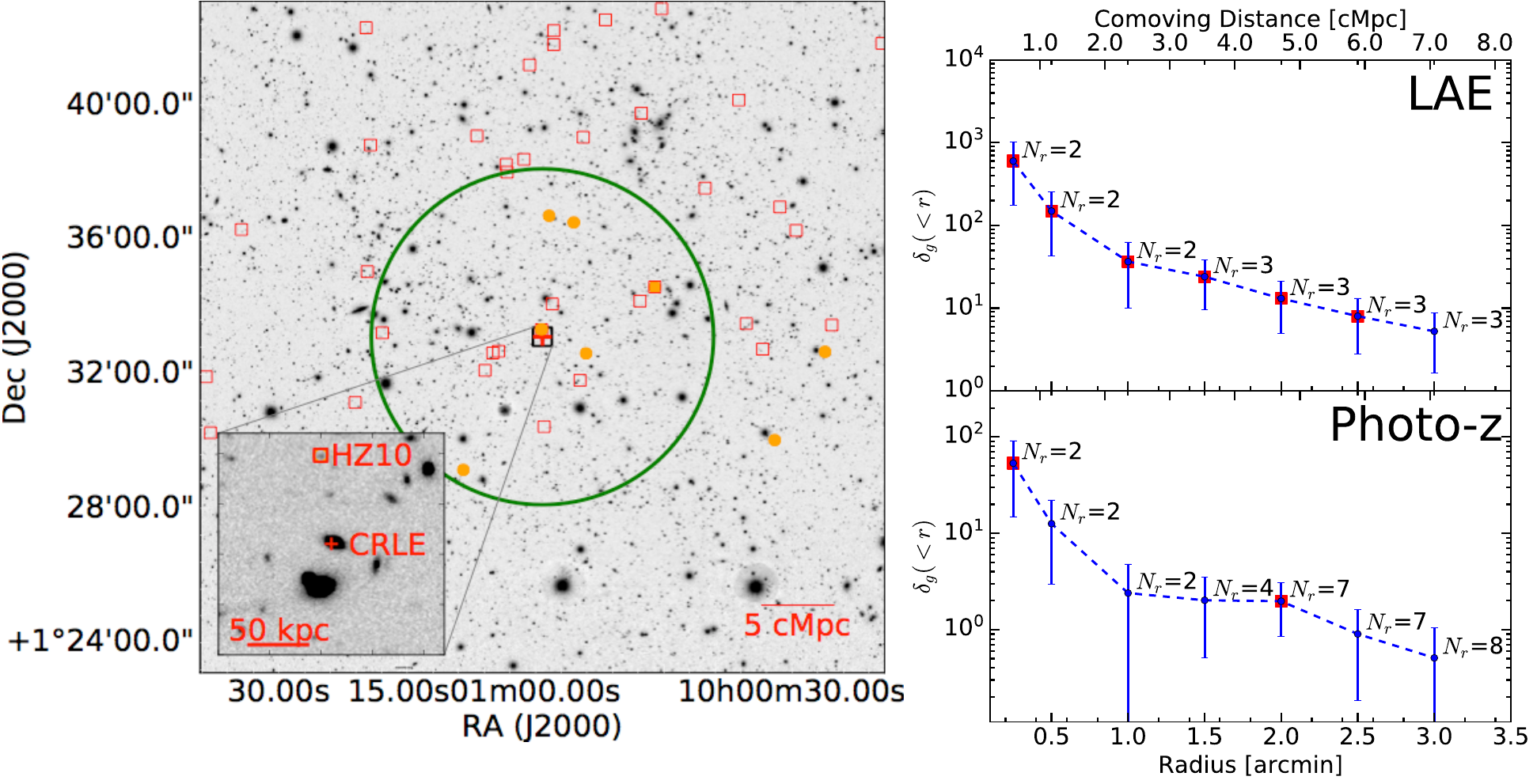}
}
\caption{Left: Galaxy overdensity around CRLE (red cross), in the two galaxy catalogs considered. The grayscale shows the narrow-band Subaru/Suprime-Cam NB816 mosaic. Orange dots represent the position of Lyman Alpha Emitters (LAE; \citealt{Murayama07}), and red squares represent galaxies from the  photometric redshift catalog at the same redshift  as CRLE \citep{Laigle16}.  The radius of the green circle is 5$^\prime$, which corresponds to $\sim12$ cMpc. The inset shows the central  $33^{\prime\prime}$ of the image, showing the relative positions of CRLE (behind a foreground galaxy) and HZ10, to the north.  Right: Overdensity parameter ($\delta_g$) as a function of radius around CRLE, evaluated for the LAE (top)  and the photometric redshift catalogs (bottom) following \cite{Vernesa_overdens}. We show the cumulative number of galaxies within the specified radius.  Red squares indicate a significant overdensity, with a false positive rate below 5\%.}
\label{fig:Overdensity}
\end{figure*}

The close association in the sky and in redshift space to HZ10 ($z=5.654$, $13^{\prime\prime}$ distance, corresponding to only 77 kpc, in projection and $\sim580$ km s$^{-1}$ of relative radial velocity) suggests that CRLE may be located in an overdense galaxy environment.
We follow the procedure described in \cite{Vernesa_overdens} to evaluate a potential galaxy overdensity around CRLE.
First, we analyze the small scale overdensity, around the DSFG in the COSMOS2015 catalog, making use  of the photometric redshift information, as detailed in \cite{Vernesa_overdens}.
Then, we apply a similar technique to also investigate the galaxy overdensity in the Lyman Alpha Emitter (LAE) catalog at $z\sim5.7$, which may offer a higher accuracy in the redshift range selected \citep{Murayama07}.

We do not apply the magnitude cut $i^+<$25.5 of \cite{Vernesa_overdens}, because the $i^+$ magnitude of HZ10 is 26.45, which we know to have a redshift in the correct range. Based on the magnitude of HZ10, we also expect that even massive galaxies at this high redshift are likely to have an $i^+$ magnitude below the adopted threshold of \cite{Vernesa_overdens}. On the other hand, our choice to include fainter galaxies may limit the photometric redshift accuracy.
We adopt their choice of redshift binning, which implies a photo-$z$ range of $\Delta z=0.64$ for the catalog ``slice".
Following \cite{Vernesa_overdens} we define a galaxy overdensity parameter by $\delta_g(r)=\frac{N_r}{\Sigma_{\rm bg}\,\pi r^2}-1$, where $N_r$ is the number of galaxies in the catalog within a radius {\it r} of the DSFG (including the DSFG, itself), and $\Sigma_{\rm bg}$ is the mean galaxy surface density in the redshift ``slice". We take into account masked regions when evaluating the surface area.
We evaluate the overdensity parameter around CRLE, at radii of 0.25$^\prime$ (to capture the overdensity due to HZ10 alone) and 0.5$^\prime$ to 3$^\prime$, in steps of 0.5$^\prime$ in both the photometric redshift  and  LAE catalogs (Figure~\ref{fig:Overdensity}).

In order to evaluate the significance of the measured overdensity parameter values, we follow the procedure by \cite{Vernesa_overdens},  producing 10 mock random catalogs adopting the same masked regions with the same number of galaxies as the real ones, and measuring the overdensity around 1000 randomly placed centers. The rate of chance occurrence of the actual  $N_r$ profile (equivalently, of the overdensity parameter) can then be evaluated. We mark significantly over-dense radii by  square markers in Figure~\ref{fig:Overdensity}, adopting the same 5\% false-positive rate as  \cite{Vernesa_overdens}.

In the COSMOS2015 photometric redshift catalog, we find that  HZ10, due to its very close separation, constitutes a significant overdensity of $>50$, and that at a radius of 2$^\prime$, there is only a 1\% probability  of the observed overdensity (7 galaxies; corresponding to $\delta_g\sim2-3$) to be produced by chance.  The chance probability of at least one neighbor within 0.5$^\prime$ is only 12\%  (the nearest, HZ10, is at 0.216$^\prime$). This chance probability is comparable to that of having at least two neighbors within 1.05$^\prime$ (i.e., the separation to the next galaxy) or of having at least three neighbors within 1.5$^\prime$ (the following galaxy).
In the LAE catalog, the overdensity is even more significant,  due to the lower contamination from galaxies at incorrect redshifts. We find a significant overdensity at all radii up to 2.5$^\prime$, with HZ10 already implying an overdensity of $>500$ and, a second LAE within 1.5$^\prime$ constituting an overdensity of $\sim30$. The overdensity in the LAE catalog at a radius of 5$^\prime$, including a total of 7 galaxies, is also $>4$. 

We also explore the second technique   utilized by \cite{Vernesa_overdens} in order to study the larger scale overdensity. However, we are limited to scales of 4--5$^\prime$ in the LAE catalog and 3$^\prime$ in the photometric redshift catalog since CRLE is close to the  southern edge of the catalog. This method utilizes a Voronoi tessellation analysis (VTA) in order to identify over-dense regions, and then determines an overdensity center by evaluating a barycenter for the ``region". It then evaluates the significance of the overdensity parameter value as a function of radius around this newly found center.
The VTA analysis in the case of CRLE indicates that, in both the photometric redshift and LAE catalogs,  CRLE and HZ10 are in an over-dense region (i.e., above the 80th percentile of a randomized galaxy density distribution), due to the close proximity to the next galaxy neighbors. The overdensity center is evaluated to be approximately the mid-point between CRLE and HZ10. Because the overdensity center is close to CRLE, the overdensity evaluation for this method reproduces similar  results to those of our previous analysis. 

Four galaxies (all part of the LAE catalog) within 5$^\prime$ have a spectroscopic redshift that place them in the overdensity (5.665,  5.674,  5.688,  and HZ10 at 5.659 from Keck/DEIMOS, Capak et al., in prep.). The other two LAEs within this radius, have no known spectroscopic redshifts\footnote{One of them, the closest LAE after HZ10, has an incorrect photometric redshift of 0.7688, the other one is not in the photometric catalog.}. Of the four LAEs with spectroscopic redshift confirmation, two (including HZ10) are also part of the photometric redshift sample. No other photometric redshift candidates within 5$^\prime$ have a spectroscopic redshift. If we expand the search radius to 17$^\prime$, corresponding to 40 cMpc in the spectroscopic redshift catalog, we find 4 more galaxies with spectroscopic redshift between 5.5 and 6, which may potentially be part of the same overdensity (with spectroscopic redshifts of 5.682,  5.728,  5.663,  5.742).



\subsection{Discussion of the overdensity significance }

Here we  compare the galaxy overdensity  around CRLE, with the cases of AzTEC-3 and other lower-redshift DSFGs  (e.g., \citealt{Walter12,Vernesa_overdens,Oteo17}), with LAE overdensities  at $z>5$ \citep{Higuchi18}, and in relation to theoretical expectations (e.g., \citealt{Chiang2013,Chiang2017}).

The closest analog to the galaxy overdensity around CRLE is the   proto-cluster around AzTEC-3 at $z=5.3$, which is also located in the COSMOS field \citep{Capak11,Riechers10a,Riechers14}. The close proximity of the ``normal" LAE/LBG  HZ10 to CRLE ($\sim77$ kpc away) is comparable to the close separation between the luminous DSFG AzTEC-3 and the ``normal" galaxy LBG-1 ($\sim95$ kpc away; \citealt{Riechers14}). \cite{Capak11} report a rich galaxy overdensity around AzTEC-3 with an overdensity parameter of $\gtrsim 11$ within a 2 cMpc radius. We also detect a comparable overdensity of LAEs of $\delta_g\gtrsim 10$ within a radius of $\sim5$ cMpc of CRLE. Because of the higher redshift of CRLE ($z\sim5.667$) than AzTEC-3 ($z\sim5.298$), catalogs contain significantly fewer galaxies. The overdensity is therefore associated with a smaller number of galaxies, and hence, is subject to larger uncertainties from random fluctuations.
The ``normal", ``companion" galaxies LBG-1 and HZ10  are similar in terms of their stellar masses and UV luminosities, but the FIR luminosity in HZ10 is almost an order of magnitude higher, suggesting that the ISM in HZ10 may by substantially more enriched (\citealt{C15,Pavesi16}). Furthermore, the [C{\sc ii}]/[N{\sc ii}] ratio appears to suggest that the  metallicity of HZ10 is compatible with local and high redshift starburst galaxies, while it suggests a lower metallicity for LBG-1 \citep{Pavesi16}.
The overdense environment experienced by HZ10 may be partly responsible for the particularly enriched ISM state in this ``normal" galaxy. In particular, it is possible that metal enriched material that was ejected by CRLE may have been accreted by HZ10. Also, it is likely that the local dark matter overdensity, as suggested by the galaxy overdensity, may be responsible for the early galaxy growth, by providing abundant gas fueling to the central regions.
However, LBG-1 seems to be located in a similar environment and does not display a comparable level of enrichment. 
\cite{Capak11} estimate that the dynamical time for LBG-1 to reach AzTEC-3 may be of order $\sim0.5$ Gyr, providing several dynamical times for a merger to occur by $z\sim2$.  A similar estimate applies to HZ10 and its likely eventual merger with CRLE, which may produce a central cluster galaxy.

\cite{Vernesa_overdens} analyzed galaxy overdensities around previously known DSFGs in the COSMOS field at $z\sim1-5.3$ and found an incidence of approximately $\sim50\%$ when using similar methods to the ones employed here. They tentatively found a higher occurrence of DSFGs occupying overdense environments at $z>3$, than at $z<3$. This would be compatible with our finding of a high galaxy overdensity including CRLE at $z\sim5.7$.
 A higher incidence of overdensities associated with the highest redshift DSFGs would be consistent with the idea that these massive galaxies may be associated with the highest peaks of the density field, tracing the most massive dark matter haloes at early cosmic epochs (e.g., \citealt{Springel05,Li07,Overzier09,Capak11}). 
 
Recent observations with the Hyper-Suprime Cam on the  Subaru telescope have yielded a deeper and wider catalog of LAEs in COSMOS at $z\sim5.7$ than available for our analysis \citep{Ouchi17,Shibuya17}.  These authors  report numerous LAE overdensities at this redshift, showing that a significant fraction of star forming galaxies at this epoch may be part of  proto-cluster environments \citep{Higuchi18}.
In particular, these authors report an overdensity,  HSC-z6PCC5, in the COSMOS field, with an overdensity parameter of $\delta\sim10$. The reported overdensity center is located 44 cMpc away from CRLE, and at a  redshift of $z=5.686$, corresponding to only 9 cMpc of radial separation. It is therefore possible that CRLE, and its associated small scale galaxy overdensity, may also be associated with this significantly larger scale early cosmic structure.

Recent theoretical work suggests that proto-clusters  may have dominated star formation in the first two billion years of cosmic time \citep{Chiang2013,Chiang2017}. This is due to the fact that  the fraction of the cosmic volume occupied by all future (proto)clusters increases by nearly three orders of magnitude from $z=0$ to $z = 7$. More importantly, most models suggest that early galaxy formation may be  dominated by the central regions of the most massive overdensities, and that star formation may evolve inside-out to galaxies in lower density environments \citep{Chiang2017}.
These may be crucial predictions of structure formation in the early Universe. A quantification of the fraction of star formation in different environments as a function of cosmic time may be an important cosmological probe in the era of wide-area surveys. The physical processes associated with the ``central"  galaxy forming in a proto-cluster, which may be a DSFG at least during part of its life, may strongly affect the evolution of such proto-clusters, both by enriching the intra-cluster medium and by providing energy input through winds and radiation.





\section{Conclusions}
We have reported the serendipitous discovery of the bright, dusty, starbursting galaxy CRLE at $z=5.667$ in the first billion years of cosmic time. This galaxy represents the highest redshift and brightest DSFG in the COSMOS field known to date, providing a higher redshift and brighter analog to the $z=5.3$ massive starburst AzTEC-3. 
We report the detection of [C{\sc ii}], [N{\sc ii}], and CO(2--1) line emission, and we find properties that are common among the highest redshift DSFGs. CRLE displays a large molecular gas reservoir ($\sim7\times10^{10}\,M_\odot$), a short gas depletion time scale of order $\sim50\,$Myr characterizing the intense starburst, and a high-intensity radiation field, as evidenced by a deep [C{\sc ii}] deficit. We find evidence for a significant fraction of the [C{\sc ii}] emission to be coming from ionized gas, similar to other high-redshift DSFGs. We suggest that this emission may be coming from a diffuse ionized medium not directly associated with the dense star-forming gas. We find dynamical evidence and dust emission properties consistent with an intermediate-stage merger. The physical proximity of the previously known ``normal" Lyman Alpha Emitter (LAE) HZ10 to CRLE constitutes a  high overdensity, and suggests that these two galaxies may coalesce in the future, forming a massive central cluster galaxy.  We find further evidence for a galaxy overdensity,  using both photometric redshift  and  LAE catalogs, which indicates the location of a likely proto-cluster analogous to the case of AzTEC-3. The presence of this likely proto-cluster supports the idea that such bright, extremely early starburst galaxies may commonly be associated with the most massive dark matter halos in the Universe at their respective epochs, providing the earliest sites of star formation  of the most massive central cluster galaxies that we observe in the local Universe.

\smallskip
\textbf{Acknowledgments}
We thank the anonymous referee for a helpful and constructive report.
RP and DR acknowledge support from the National Science Foundation under grant number
AST-1614213 to Cornell University.
RP acknowledges support through award SOSPA3-008 from the NRAO. 
The National Radio Astronomy Observatory is a facility of the National Science Foundation operated under cooperative agreement by Associated Universities, Inc.
This paper makes use of the following ALMA data: ADS/JAO.ALMA\#2015.1.00928.S, 2015.1.00388.S, 2012.1.00523.S. ALMA is a partnership of ESO (representing its member states), NSF (USA) and NINS (Japan), together with NRC (Canada), MOST and ASIAA (Taiwan), and KASI (Republic of Korea), in cooperation with the Republic of Chile. The Joint ALMA Observatory is operated by ESO, AUI/NRAO and NAOJ.

\appendix

\section{A. Constraining the effects of gravitational Lensing}


  The high apparent luminosity of CRLE, combined with the coincident spatial position with a foreground galaxy, raise questions concerning the relative importance of strong gravitational lensing. About 10\% of the area of the segmentation map in the {\em HST} H-band images we utilized to study CRLE is occupied by (mostly) local galaxies.  Therefore,  the coincidence of galaxies at different redshifts may not be uncommon.
Using the method by \cite{Harris12}, we roughly estimate the magnification due to gravitational lensing, based on the apparent correlation between intrinsic CO luminosity and line width, finding $\mu=0.9\pm 0.2$. This may suggest that lensing is not expected to significantly boost the observed luminosity of CRLE. However, we note that the \cite{Harris12} method relies on a proposed $L'_{\rm CO}$--FWHM$_{\rm CO}$ relation, which may only hold with large scatter (e.g., \citealt{Sharon16}).

We also explore the  potential lensing magnification by modeling the foreground galaxy with the commonly used approximation of a singular isothermal sphere (SIS). We confirm that the light distribution appears to be approximately proportional to the aperture radius, which implies that a ``light traces mass" model would  be approximately equivalent to the adopted SIS. 
By measuring the flux in the {\em HST}/WFC3 F160W image in apertures of varying radius, we deduce that 68\% of light is included within  1$^{\prime\prime}$ radius, 37\% within 0.5$^{\prime\prime}$ radius, 22\% in 0.3$^{\prime\prime}$ radius. We assume that the total mass in the central regions of the foreground galaxy should be dominated by the stellar mass. Therefore, we adopt a total mass of $3.3\times10^9\,M_\odot$, as  reported by the COSMOS2015 catalog \citep{Laigle16}, in agreement with our {\it Cigale} SED modeling.
A combination of the distance implied by the photometric redshift, the measured size, and the total mass suggests a velocity dispersion parameter for the SIS model of only 25--30 km s$^{-1}$. This velocity dispersion parameter corresponds to a very small Einstein radius of $\sim$0.02$^{\prime\prime}$--0.025$^{\prime\prime}$, at the lens and source distances derived by their redshifts. 
We measure a separation of the DSFG from the lens center of $\sim$0.3$^{\prime\prime}$.  If we were to assume a point source model for the DSFG, we would estimate a magnification of $<$10\%.
The relative positional uncertainty is dominated by the {\em HST} astrometric uncertainty, which we assume to be $\sim$0.1$^{\prime\prime}$ (e.g., \citealt{Carlos_SMGs}), dominating over the ALMA positional accuracy or the fitting uncertainty. 

We use both our custom code and the publicly available code \texttt{gravlens} \citep{Keeton01} to constrain the effect of lensing in a model source distribution, obtaining equivalent answers.
We fix the lens Einstein radius to 0.025$^{\prime\prime}$, the source spatial FWHM  to 0.6$^{\prime\prime}$ based on the measured   [C{\sc ii}] size, and a mean lens-source separation of 0.3$^{\prime\prime}$.
We randomly vary this positional separation by adding independent, normally distributed spatial offsets along the two axes with standard deviation of 0.1$^{\prime\prime}$, representing the relative positional uncertainty.
This results in an approximately normally distributed magnification of $\mu=1.09\pm0.02$.
Furthermore, since the Einstein radius is so small compared to the spatial extent of the source ($<5\%$), the effects of lensing are negligible both to the global flux but also to the observed kinematic structure within the uncertainty of our measurements.
In addition, we also allow the source size and relative position to vary, and verify that a significantly smaller intrinsic source size is incompatible with the observed source size. The small Einstein radius implies that the observed source size can only be reproduced by a comparable intrinsic size, therefore significantly constraining the allowed magnification to the values reported above.
We therefore assume that lensing does not measurably affect any of our conclusions based on the luminosity and spatial structure of CRLE.

\section{B. HST foreground galaxy removal}

\begin{figure*}[htb]
\centering{
\includegraphics[width=.8\textwidth]{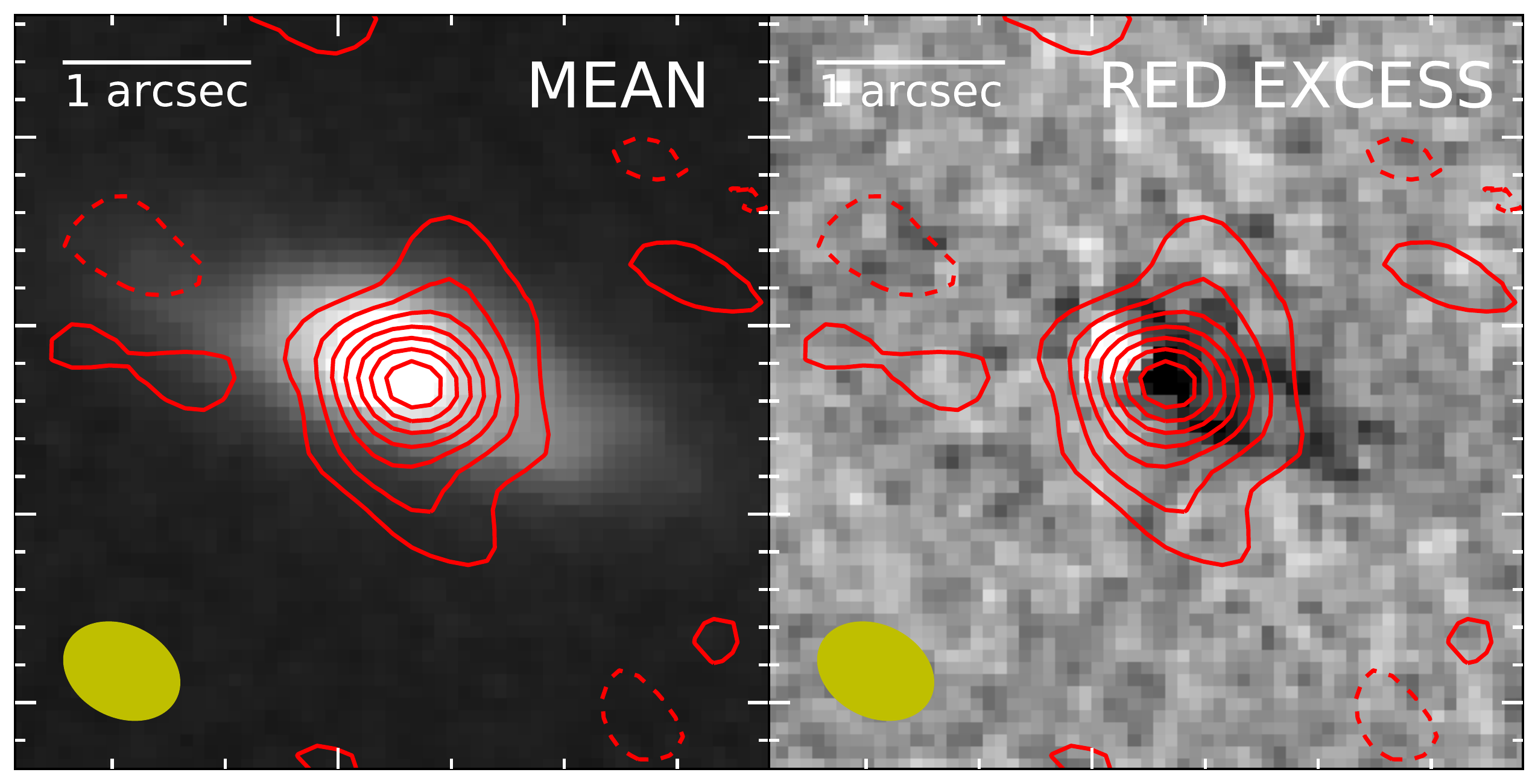} 
}
\caption{{\em HST} NIR band mean emission and red color excess, attempting to separate the emission from CRLE.
Left: ALMA [C{\sc ii}] contours shown in steps of $2\sigma$, overlaid on {\em HST} NIR mean image. To obtain the mean image, we averaged the emission detected in the three {\em HST}/WFC3 bands F105W, F125W and F160W. Right: Difference map of F160W and the mean image, showing F160W emission in excess from the mean.}
\label{fig:red_excess}
\end{figure*}

\begin{figure*}[htb]
\centering{
  \includegraphics[width=.93\textwidth]{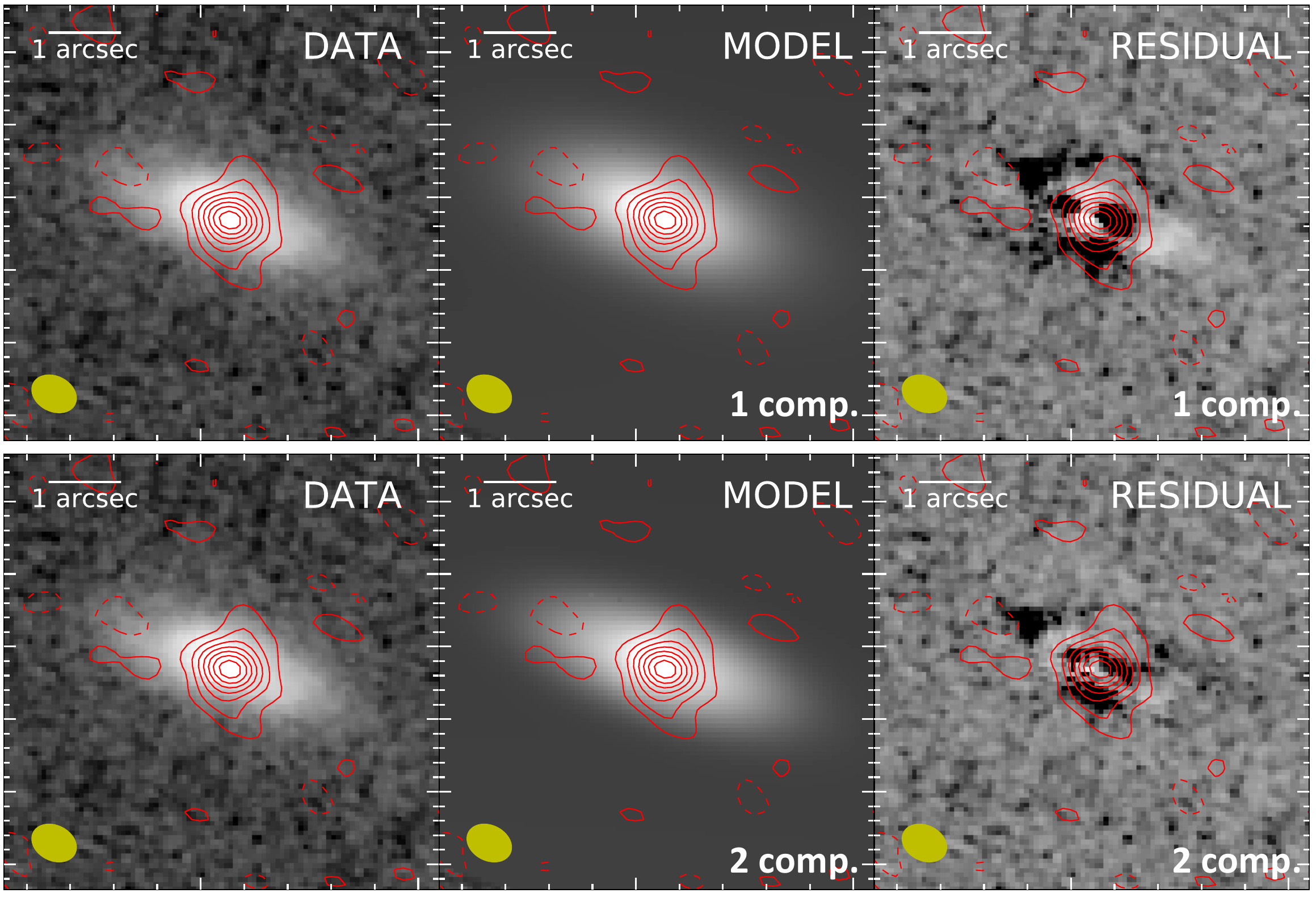} 
}
\caption{Top: Single S\'ersic component model fit with {\it Galfit} to the {\em HST}/WFC3 F160W emission from the foreground galaxy. The low level residuals show a hint of emission (a few percent in flux) which may be associated with asymmetric structure in the foreground galaxy or CRLE itself (position indicated by the ALMA [C{\sc ii}] contours). Bottom: Two-component {\it Galfit} best-fit model and residuals, showing that no significant additional structure is present.
 }
\label{fig:galfit}
\end{figure*}

In order to constrain the rest-frame optical emission from CRLE, we attempt different methods for removing the contamination of the {\em HST}/WFC3 NIR images due to the foreground galaxy. First, we attempt to separate the emission from CRLE from that of the foreground galaxy based on a color difference between the two galaxies (Figure~\ref{fig:red_excess}). The mean NIR image (average of WFC3/F105W, F125W, and F160W) shows smooth emission due to the foreground disk galaxy  but a F160W ``excess" (relative to the mean NIR emission) shows a more red than average component to the northeast of the central position (corresponding to only $\sim0.5\%$ of the total F160W flux) and a deficit of F160W emission at the position of the CRLE [C{\sc ii}] emission peak. We may expect the rest-frame optical emission associated with CRLE to appear redder than that of the foreground galaxy.  This tentative evidence may therefore suggest that the stellar emission from CRLE may be offset from the [C{\sc ii}]  peak as frequently observed in high redshift DSFGs perhaps due to differential dust obscuration or an older stellar population offset from the young massive star-forming regions (e.g., \citealt{Riechers14,Carlos_SMGs}). We also use {\it Galfit} to fit S\'ersic profile emission models and remove the foreground emission from the F160W image \citep{Galfit}. We first fit a single component model characterized by the center position, flux, radius, S\'ersic index, axis ratio and position angle (Figure~\ref{fig:galfit}, top). We find a S\'ersic index of $\sim1$, compatible with an exponential disk, and a half-light radius of 0.82$^{\prime\prime}$. The total emission is not fit well by this model and shows positive residuals to the northeast of the center, which may or may not be associated with the ``red" excess seen in Figure~\ref{fig:red_excess}. The flux associated with this positive residual is only $\sim1-2\%$ of the total F160W flux. Since part of the foreground galaxy emission to the west is apparent in the residuals, we do not consider these residuals to be sufficient to indicate an additional source of emission, beyond the imperfect model fit of an intrinsically not perfectly symmetric galaxy. In order to achieve a better fit to the foreground emission we also fit a two-component S\'ersic profile (Figure~\ref{fig:galfit}, bottom). The residuals do not indicate significant additional structure. The second emission component may be associated with structure in the foreground galaxy or with emission from CRLE. In the main text, we  neglect the contribution from CRLE to the emission at these wavelengths because it appears to represent, at most, a few percent of the total.

\section{C. Dynamical modeling}

\begin{figure}[htb]
\centering{
  \includegraphics[width=.48\textwidth]{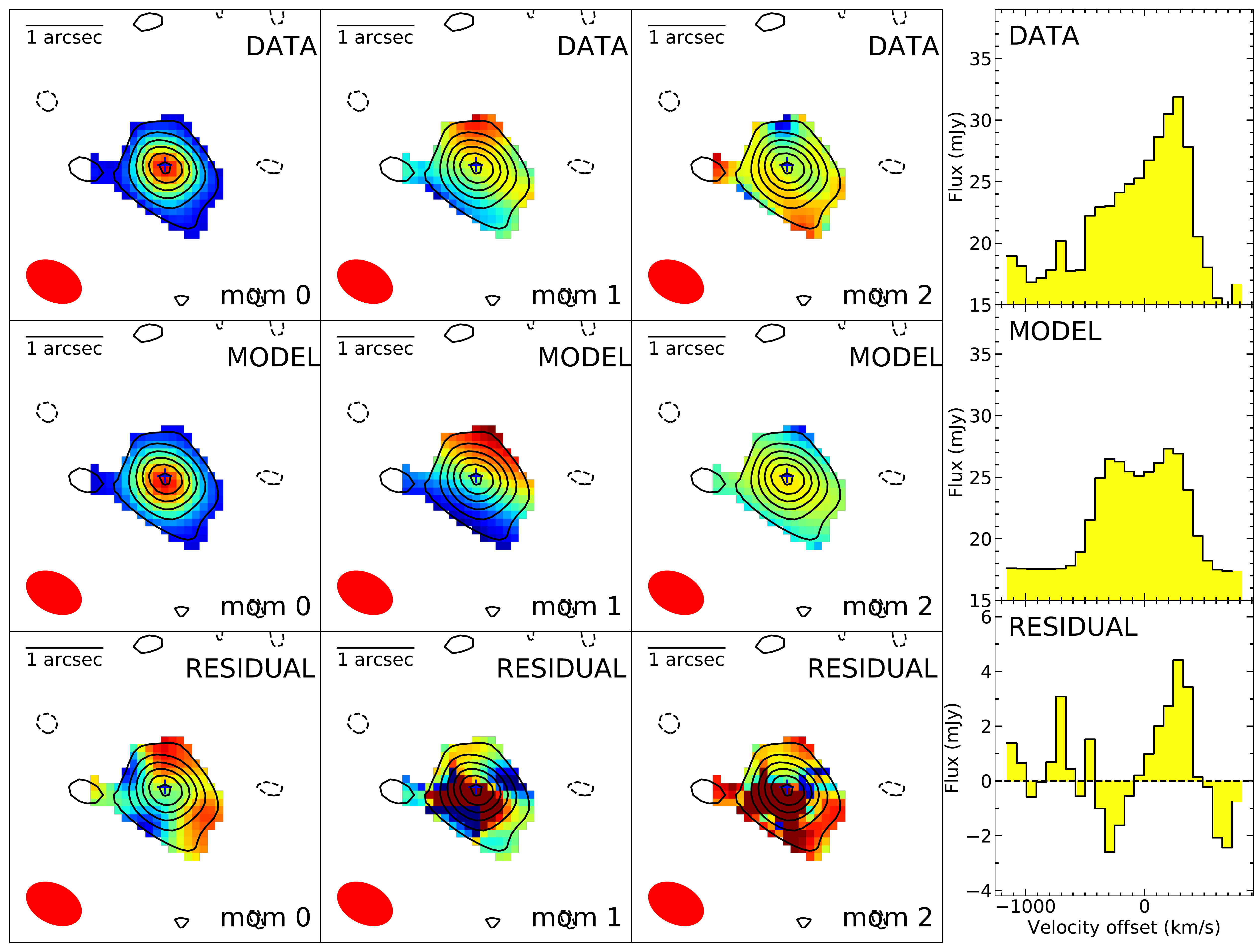} \includegraphics[width=.48\textwidth]{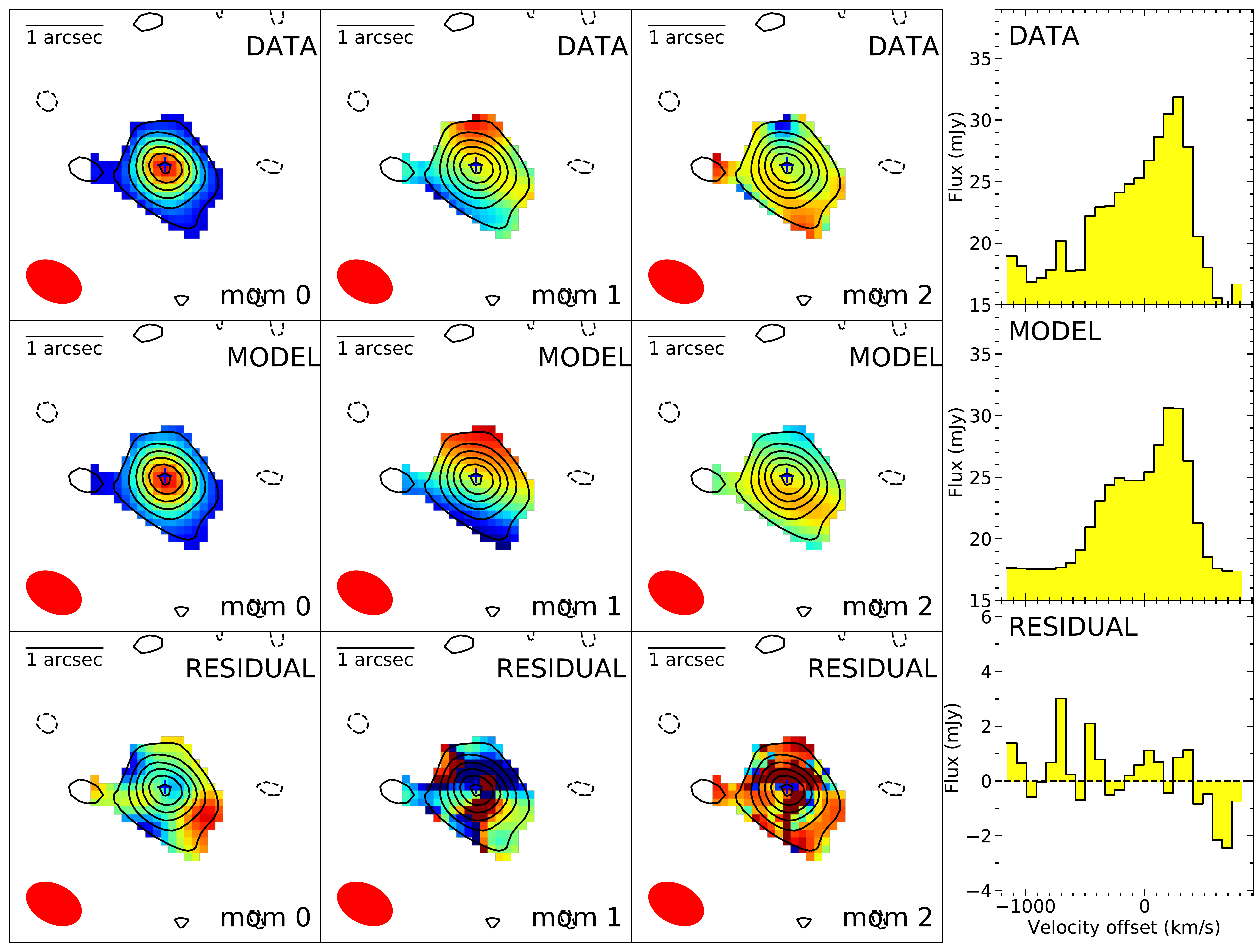}

}
\caption{Visibility space dynamical modeling results for the [C{\sc ii}] line emission in CRLE. We show the ``natural" weighting line moment 0 (intensity), 1 (velocity) and 2 (dispersion) maps and spectra for the data, the model, and the visibility residuals.
Left: One component rotating disk model. Right: Two component model, with a spatially unresolved second dynamical component. Although a one component model provides an acceptable fit, the residuals show clear spectral structure, and the moment-2 map shows spatial structure hinting to a different dynamical component to the north. The residuals are significantly improved by the addition of a second model component, perhaps suggesting an ongoing galaxy merger.}
\label{fig:dyn_model}
\end{figure}

\begin{table}[t]
\centering{
\caption[]{Results of one- and two-component dynamical modeling.}
\label{table_1comp_dyn}
\begin{tabular}{c|ccc|ccc}
\hline 
&&One component&&&Two component&\\
Parameter (Units)&16th perc.&50th perc.&84th perc.&16th perc.&50th perc.&84th perc.  \\
\hline \noalign {\smallskip}
Gas dispersion (km s$^{-1}$)&98&105&112&118&129&141\\
Emission FWHM (arcsec)&0.43&0.45&0.47&0.68&0.71&0.75\\
Maximum velocity (km s$^{-1}$)&640&800&1100&413&435&470\\
Velocity scale length (arcsec)&0.010& 0.012&  0.015&0.014&0.026&0.053\\
Inclination (degrees)&20&28&36&57&60&64\\
Position angle (degrees)&110&114&117&93&96&99\\
\hline
Continuum major FWHM (arcsec)&0.42&0.43&0.44&0.42&0.43&0.44\\
Continuum minor FWHM (arcsec)&0.26&0.27&0.29&0.26&0.27&0.29\\
\hline
Second component velocity FWHM (km s$^{-1}$)&---&---&---&250&270&290\\
\hline \noalign {\smallskip}
\end{tabular}
}
\end{table}

In order to extract the most precise physical parameters for CRLE, we analyze the [C{\sc ii}] line data, which have the highest signal-to-noise ratio and angular resolution,  by dynamical model fitting.

We have developed a novel implementation of dynamical model fitting, working directly on the visibilities. Carrying out the model comparison to data in the {\sc uv} space, rather than in the image plane, makes our method independent of deconvolution and of choice of visibility weighting, and hence more robust. Furthermore, our method takes advantage of the well-behaved (i.e., independent and normally distributed) noise properties of measured visibilities, in comparison to the correlated noise affecting interferometric images.

Our method has general applicability for interferometric data with frequency structure, and is based on a  Bayesian formulation of the model fitting problem. We use Markov Chain Monte Carlo (MCMC) and Nested Sampling techniques in the form of {\it emcee} \citep{emcee} and {\it MultiNest} for \texttt{python} \citep{Multinest,pymultinest} to sample the posterior distribution for the model parameters, and to evaluate the model evidence (also called marginal likelihood), i.e., the integral of the posterior which gives the probability of the model given the data. We have verified that the parameter estimates derived from the samples produced by the two different techniques are well within the range of compatibility.

The first model  fit to  CRLE [C{\sc ii}] is a rotating disk model, generated through the publicly available code {\it KinMSpy} \citep{KinMS}, superposed onto an elliptical 2D Gaussian continuum model.  We choose this model to be described by disk center coordinates, systemic velocity, gas dispersion, FWHM size of the spatial light profile of the disk (assumed to be 2D Gaussian), maximum velocity and velocity scale length, inclination, position angle and line flux.
The continuum flux and FWHM sizes of the continuum emission were separately fixed by fitting the line-free channels.
We impose non-constraining priors. We choose uniform priors for additive parameters, logarithmic priors for scale parameters, and a sine prior for the inclination angle. The $1\sigma$ confidence intervals of the physically relevant parameters derived from our modeling are shown in Table~\ref{table_1comp_dyn}, and the median model fit and the residuals are shown in Fig.~\ref{fig:dyn_model}. 

Clear structure is visible in the single component model residuals by adopting median parameters, particularly in the spectrum, although the total residual flux has formally a low significance (Fig.~\ref{fig:dyn_model}). We explore a second model, with five additional parameters describing a second, unresolved component, which is designed to capture the narrow velocity  component visible in the spectra and in the dispersion map. 
The additional parameters describe the second component center x and y coordinates, systemic velocity, integrated flux and line velocity width. 
As Fig.~\ref{fig:dyn_model} shows, the model fit is significantly improved by this additional component. We characterize the improvement to the quality of the model fit achieved by the addition of the second component by the model evidence ratio evaluated through {\it MultiNest} of $e^{86}\sim10^{37}$, which takes into account the additional parameter space available to the more general model. 
Therefore, we conclude that a single component rotating disk model is not sufficient to describe the [C{\sc ii}] emission in CRLE, and we find, instead, strong indication for a second component corresponding to the narrow line emission component also observed in the CO and [N{\sc ii}] lines.
Higher resolution observations are required in order to determine if coherent rotation may be important to the gas dynamics in this system, or whether the dynamics are dispersion dominated. The latter would provide stronger evidence in favor of a merging pair of galaxies, perhaps identified by the two separate dynamical components. However, strong shocks and winds may also be responsible for skewing the velocity profile, and could therefore contribute to the observed dynamics.





\vspace{.3in}
\bibliographystyle{aasjournal}

\bibliography{CRLE_biblio}

\end{document}